\documentclass[prd,preprint,amsmath,amssymb,nofootinbib,eqsecnum]{revtex4-1}
\usepackage{latexsym,graphics,color,epsfig,hyperref,mathrsfs}

\newcommand{\ie}{{i.e.}}
\newcommand{\cf}{{cf.}}

\newcommand{\wrt}{with respect to}
\newcommand{\lhs}{left-hand side}

\newcommand{\rhs}{right-hand side}

\newcommand{\naive}{na\"{\i}ve}

\newcommand{\be}{\begin{equation}}
\newcommand{\ee}{\end{equation}}
\newcommand{\bea}{\begin{eqnarray}}
\newcommand{\eea}{\end{eqnarray}}
\newcommand{\beas}{\begin{eqnarray*}}
\newcommand{\eeas}{\end{eqnarray*}}

\newcommand{\bear}{\begin{array}{l}}
\newcommand{\eear}{\end{array}}

\newcommand{\bcf}{\begin{center}\begin{figure}}
\newcommand{\ecf}{\end{figure}\end{center}}

\newcommand{\bct}{\begin{center}\begin{table}}
\newcommand{\ect}{\end{table}\end{center}}

\newcommand{\eq}[1]{(\ref{eq:#1})}

\newcommand{\eqs}[2]{(\ref{eq:#1}) and~(\ref{eq:#2})}
\newcommand{\eqss}[3]{(\ref{eq:#1}), (\ref{eq:#2}) and~(\ref{eq:#3})}

\newcommand{\sect}[1]{section~\ref{sec:#1}}

\newtheorem{prop}{Proposition}

\newtheorem{representation}{Representation}

\newcommand{\D}{d}
\newcommand{\Int}[1]{\int \!\! d^{\D} \! #1 \,}
\newcommand{\volume}[1]{d^{\D} \! #1 \,}

\newcommand{\DD}[1]{\delta^{(#1)}}

\newcommand{\identity}{\mathbb{I}}

\newcommand{\pder}[2]{\frac{\partial #1}{\partial #2}}

\newcommand{\fder}[2]{\frac{\delta #1}{\delta #2}}

\newcommand{\Or}{\mathrm{O}}
\newcommand{\order}[1]{\Or \bigl( #1 \bigr)}

\newcommand{\hf}{\frac{1}{2}}

\newcommand{\tr}{\mathrm{tr}\,}

\newcommand{\eval}[1]{
	\bigl\langle #1 \bigr\rangle
}

\newcommand{\one}{1\!\mathrm{l}}

\newcommand{\Rot}{\mathcal{M}}
\newcommand{\Trans}{\mathcal{P}}
\newcommand{\SCT}{\mathcal{K}}
\newcommand{\Dil}{\mathcal{D}}

\newcommand{\RotR}{\mathscr{M}}
\newcommand{\TransR}{\mathscr{P}}
\newcommand{\SCTR}{\mathscr{K}}
\newcommand{\DilR}{\mathscr{D}}

\newcommand{\rot}[1]{L_{#1}}
\newcommand{\dil}[1]{D^{(#1)}}
\newcommand{\sct}[2]{{K^{(#1)}}_{#2}}

\newcommand{\dilL}[1]{\overleftarrow{D}^{(#1)}}
\newcommand{\sctL}[2]{{\overleftarrow{K}^{(#1)}}_{#2}}

\newcommand{\operator}{\mathcal{O}}
\newcommand{\quasi}{\mathscr{O}}
\newcommand{\qsource}{J}
\newcommand{\scaling}{\Delta}

\newcommand{\dfield}{\varphi}

\newcommand{\cutoff}{K}

\newcommand{\ep}{\mathcal{G}}
\newcommand{\op}{\mathcal{Y}}

\newcommand{\Stotd}{\mathcal{S}^{\mathrm{tot}}}
\newcommand{\Sintd}{\mathcal{S}}

\newcommand{\dual}{\mathcal{W}}

\newcommand{\hepth}[1]{hep-th/#1}
\newcommand{\hepph}[1]{hep-ph/#1}

\begin{document}

\title{On Functional Representations of the Conformal Algebra}
\author{Oliver J.~Rosten}
\affiliation{oliver.rosten@gmail.com}
\begin{abstract}
	Starting with conformally covariant correlation functions, a sequence of functional representations of the conformal algebra is constructed. A key step is the introduction of representations which involve an auxiliary functional. It is observed that these functionals are not arbitrary but rather must satisfy a pair of consistency equations corresponding to dilatation and special conformal invariance. In a particular representation, the former corresponds to the canonical form of the Exact Renormalization Group equation specialized to a fixed-point whereas the latter is new. This provides a concrete understanding of how conformal invariance is realized as a property of the Wilsonian effective action and the relationship to action-free formulations of conformal field theory.
	
Subsequently, it is argued that the conformal Ward Identities serve to define a particular representation of the energy-momentum tensor. Consistency of this construction implies Polchinski's conditions for improving the energy-momentum tensor of a conformal field theory such that it is traceless.
In the Wilsonian approach, the exactly marginal, redundant field which generates lines of physically equivalent fixed-points is identified as the trace of the energy-momentum tensor. 

\vspace{10ex}
\begin{center}
\large{In loving memory of Francis A.~Dolan.}
\end{center}
	
\end{abstract}

\maketitle
\tableofcontents

\section{Introduction}

\subsection{Conformal Field Theories}

The essential information content of any quantum field theory (QFT) is encoded in its correlation functions. As such, the various different approaches to the former ultimately amount to different strategies for computing the latter. In an ideal situation, it could be imagined that the correlation functions are determined entirely by some symmetry, allowing one to concentrate solely on the representation theory of the appropriate algebra, having dispensed with standard notions such as an action and corresponding path integral.

In general, such a strategy is not available. However, a partial realization occurs for QFTs exhibiting conformal symmetry---the Conformal Field Theories (CFTs). If we suppose that the correlation functions involve a set of local objects, $\{\operator(x)\}$, then a special set of `conformal primary fields', $\quasi_i(x)$, can be identified for which the correlation functions exhibit covariance under global conformal transformations. Focussing on the conformal primaries, the various two and three-point correlation functions are determined by the conformal symmetry in terms of the \emph{a priori} unknown CFT data: the scaling dimensions, $\scaling_i$, spins and three-point coefficients, $C_{ijk}$. At the four-point level and beyond, the direct constraints of conformal symmetry are weaker, still.

Further progress can be achieved by applying the Operator Product Expansion (OPE). Within correlation functions, consider taking a limit such that the positions of two of the fields approach each other. According to the OPE, in this limit the pair of fields can be replaced by a linear sum over fields; schematically, this can be written
\be
	\operator_a(x) \operator_b(y) \sim \sum_c f_{ab}^c(x-y) \operator_c(y)
.
\ee
A particularly powerful effect of conformal symmetry is that the complete content of the OPE can be rephrased in terms of just the conformal primary fields. If the OPE converges for finite separations, then $n$-point correlation functions can be determined in terms of $n-1$ point correlation functions. In this manner, the content of conformal field theories can, in principle, be boiled down to the CFT data introduced above.

However, to determine the various combinations of CFT data which correspond to actual CFTs (possibly subject to constraints, such as unitarity) requires further input. One approach is to exploit associativity of the OPE to attempt to constrain the CFT data (this technique is known as the conformal bootstrap). In general, the task is extremely challenging since one can expect an infinite number of conformal primaries, the scaling dimensions of which must be self-consistently determined. Nevertheless, substantially inspired by work of Dolan and Osborn~\cite{Francis-4ptOPE,Francis-PartialWavesOPE,Francis-FurtherResults}, remarkable recent progress has been made in this area~\cite{Rychkov-4DBounds,Rychkov-SpinningBlocks,Poland-4DBounds,Rychkov-IsingI,Rychkov-IsingII,Gliozzi:2013ysa,Gliozzi:2014jsa}.

In two dimensions, additional structure is present. Whilst the global conformal group is always finite dimensional, in $\D=2$ there exists an infinite dimensional local conformal algebra, the Virasoro algebra. Fields can now be classified according to their transformations under local (rather than just global) conformal transformations and, as such, arrange themselves into multiplets comprising a Virasoro primary and its descendants. In the seminal paper~\cite{BPZ} it was shown that there are a set of special theories, the `minimal models' for which there are only a finite number of Virasoro primaries possessing known scaling dimensions. This simplification is sufficient for the bootstrap procedure to determine the CFT data and for all correlation functions to be expressible as solutions to linear partial differential equations.

However, even in this situation, there are some natural questions to pose, at least coming from the perspective of the path integral approach to QFT: is it possible to encode the dynamics of these theories in an action and, if so, is there a concrete recipe for doing so? Is this procedure possible for all such theories or only some of them? Are the resulting actions guaranteed to be local and, if so, why? In this paper, it will be attempted to provide answers to some of these questions and hopefully to offer a fresh perspective on others.

These questions are equally valid (and perhaps less academic) in situations where the conformal bootstrap is insufficient to provide a complete understanding of the theories to which it is applied. We will take the point of view that, in this situation, one 
approach is to try to introduce an action formulation of the theory in question. Notice that this has been deliberately phrased so as to reverse the logical order compared to the path integral approach. Typically, within the path integral paradigm, the first thing that one does is write down a (bare) action. The conformality or otherwise of the resulting theory must then be determined. In our approach, however, we envisage starting with correlation functions which are conformally covariant by assumption and then introducing the action as an auxiliary construction.

At heart, the underlying philosophy of this paper is to take an intrinsically quantum field theoretic starting point i.e.\ the symmetry properties of the correlation functions of the theory at hand. The idea then is to show that, perhaps given certain restrictions, this \emph{implies} that (should one so desire) a local action can be constructed, from which the correlation functions can, in principle, be computed. If one is to take QFT as fundamental, this seems more philosophically satisfactory than taking a classical action as the starting point. Moreover, it clarifies the relationship between two largely disassociated textbook approaches to QFT. 

\subsection{The Exact Renormalization Group}
\label{sec:ERG-intro}

As will be exposed in this paper, the formalism which binds together the classic CFT approach to field theory with its path integral counterpart is Wilson's Exact Renormalization Group (ERG)~\cite{Wilson}. Starting from conformally covariant correlation functions, the strategy is to encode the information thus contained in various functional representations.%
\footnote{While the correlation functions themselves are conformally \emph{co}variant, we refer to the functionals as representing a conformally \emph{in}variant theory.
}
Each representation will yield different expressions both for the conformal generators and for the conformal primaries and their descendants.
The most direct representation follows from introducing sources and embedding the correlation functions in the Schwinger functional, $\dual$ (the subtleties of doing this in the presence of infrared (IR) and ultraviolet (UV) divergences are discussed later). Associated with the Schwinger functional is a representation of the conformal algebra; for CFTs, each of the generators annihilates $\dual$.

For simplicity, we will largely consider theories for which all conformal primary fields are scalar (relaxing this, at least in the absence of gauge symmetry, is straightforward). More importantly, we assume that there is at least one conformal primary field for which the Schwinger functional---written in terms of the conjugate source, $\qsource$---exists. To proceed, we shift this source by the derivative of a new, auxiliary field:
\be
	\qsource \rightarrow \qsource + \partial^2 \dfield,
\label{eq:J-shift}
\ee
resulting in another representation of the conformal algebra. Note that $\dfield$ will essentially end up playing the role of the fundamental field (for brevity, we implicitly consider theories which involve just a single fundamental field).

The form of the shift~\eq{J-shift} may seem a little odd. Ultimately, it can be traced back to Morris' observation that the Wilsonian effective action naturally generates the correlation functions with an extra factor of momentum squared on each leg~\cite{TRM-ApproxSolns}. 

Next, we introduce a deformation of the Schwinger functional obtained by adding an apparently arbitrary functional of $\dfield$ and $\qsource$ which, amongst other things, means that the deformed functional may depend separately on the field and the source. The motivation for this follows from previous studies of the ERG: this deformed functional is recognized as something which can be generated from a Wilsonian effective action, $\Sintd$. To put it another way, we know the answer that we're looking for! 

Recall that, in the Schwinger functional representation, a CFT is such that the generators annihilate $\dual$. This is translated into the statement that the generators in the new representation annihilate $\exp{-\Sintd}$. However, whereas the generators in the Schwinger functional representation are linear in functional derivatives, in the new representation the generators associated with dilatations and special conformal transformations are quadratic. The upshot of this is that two of the linear conditions implied by the annihilation of  $\exp{-\Sintd}$ can be replaced two by non-linear conditions on $\Sintd$. The first of these is identified with the fixed-point version of an ERG equation and the second---constituting one of the central results of this paper---is recognized as a new analogue of the special conformal consistency condition discovered long ago by Sch\"afer~\cite{Schafer-Conformal}. Associated with these conditions is a representation of the conformal algebra in which the generators depend explicitly on the Wilsonian effective action. 

The non-linearity of the ERG equation seems to be crucial (as emphasised by Wegner~\cite{Wegner-CS}). In the linear Schwinger functional representation, the scaling dimensions of the fields must be determined in a self-consistent fashion using the bootstrap equations. In the Wilsonian approach, a different strategy is used. One field%
\footnote{%
In much of the ERG literature, `operator' is used in place of field; however, following CFT conventions, we shall generally use the latter.
}, with \emph{a priori} unknown scaling dimension, $\delta$, is separated from the rest and used to formulate an ERG equation (as anticipated above, we identify this field as the fundamental field). As such,  it appears that an ERG equation contains two unknowns: the Wilsonian effective action and $\delta$. The correct interpretation is that an ERG equation is a non-linear eigenvalue equation~\cite{TRM-Elements}; however, this hinges on one further ingredient: we demand that the solutions to the ERG equation are quasi-local.%
\footnote{By quasi-local it is meant that contributions to the action exhibit an expansion in positive powers of derivatives. Equivalently, in momentum space, vertices have an all-orders expansion in powers of momenta. Loosely speaking, at long distances a quasi-local action has the property that it reduces to a strictly local form. Quasi-locality is discussed, at length, in~\cite{Fundamentals}.}

It is the combination of non-linearity and quasi-locality which allows, in principle, for the spectrum of $\delta$ to be extracted from the ERG equation. Indeed, by demanding quasi-locality, the spectrum of possible values of $\delta$ can be shown to be discrete~\cite{Fundamentals}. 
Let us emphasise that the spectrum of $\delta$ does not correspond to the spectrum of fields within a given CFT; rather, each value of $\delta$ obtained by solving the ERG equation corresponds to a different CFT. 

Presuming that some solution to the ERG equation has been obtained, the second step would be to compute the spectrum of fields. With both $\Sintd$ and $\delta$ now known, the dilatation generator has a concrete form. It now provides a linear eigenvalue equation for the fields and their scaling dimensions. In general, the spectrum is rendered discrete by the condition of quasi-locality: this is illustrated for the Gaussian fixed-point in~\cite{Wegner-CS,Fundamentals}. Within the derivative expansion approximation scheme, see~\cite{TRM-Elements} for an excellent description of how discreteness of the spectrum arises for non-trivial fixed-points.

From the perspective of different representations of the conformal algebra, it is locality, together with non-linearity, which singles out the ERG representation as special. Remember that the reason for considering elaborate representations of the conformal algebra is more than just academic: part of the motivation is to provide tools for understanding conformal field theories for which the conformal bootstrap seems intractable. In the ERG approach, scale-invariant theories can be picked out from an equation by applying a constraint which is easy to implement on the solutions: that of quasi-locality.
 The price one pays for this is the introduction of a considerable amount of unphysical scaffolding, notably the UV regularization. One can imagine other representations of the conformal algebra which entail similar complication but without the redeeming feature of a simple condition which can be imposed on solutions of the scale/special conformal consistency conditions.

An interesting question to ask is whether the set of conformal field theories (perhaps subject to constraints of physicality) is in one-to-one correspondence with the set of equivalence classes of quasi-local actions. It is tempting and perhaps not too radical to speculate that, at the very least for the sorts of theories which form the focus of this paper---non-gauge theories, on a flat, static background, for which the energy-momentum tensor exists and is non-zero---the answer is yes. While there are some suggestive numerical results~\cite{TRM-2D}, it is desirable to have a proof, one way or the other. The thrust of this paper gives some clues as to how this can be achieved (as discussed further in the conclusion)
but at a rigorous level the question  remains unanswered, for now.

\subsection{The Energy-Momentum Tensor}

It is worth emphasising that the approach advocated above is precisely the opposite of the standard path integral approach: our starting point is the correlation functions; then we introduce sources; next we introduce a field and arrive at an action! Everything within this picture, with the exception of the correlation functions, themselves, is an auxiliary construction. As such, this sits rather uncomfortably with standard expositions of the role of the energy-momentum tensor in QFT: for these tend to start with the classical action.

This tension is reconciled as follows. One of the key results of this paper is that, for conformal field theories, the usual Ward identities associated with the energy-momentum tensor should, in fact, be recognized as \emph{defining} the energy-momentum tensor in a particular representation of the conformal algebra.%
\footnote{Actually, there are CFTs, such as the Mean Field Theories, for which the energy-momentum tensor does not exist; this will be discussed further in \sect{proposal}.}
 To see how this comes about, consider the Ward identity associated with translation invariance. Taking $T_{\alpha\beta}$ to denote a quasi-local representation of the energy-momentum tensor, we have~\cite{YellowPages}:
\be
	\partial_\alpha
	\eval{T_{\alpha \beta}(x) \quasi^{(\delta)}_{\ \mathrm{loc}}(x_1) 
	\cdots \quasi^{(\delta)}_{\ \mathrm{loc}}(x_n)}
	=
	-\sum_{i=1}^n
	\DD{\D}(x-x_i) \pder{}{x_{i\beta}}
	\eval{\quasi^{(\delta)}_{\ \mathrm{loc}}(x_1) \cdots \quasi^{(\delta)}_{\ \mathrm{loc}}(x_n)}
\ee
where $\quasi^{(\delta)}_{\ \mathrm{loc}}$ is a quasi-local representation of the conformal primary field conjugate to $\qsource$.
Multiplying by one source for each instance of the quasi-primary field and integrating over the corresponding coordinates yields:
\be
	\partial_\alpha
	\eval{T_{\alpha \beta}(x) \qsource \cdot \quasi^{(\delta)}_{\ \mathrm{loc}} 
	\cdots  \qsource \cdot \quasi^{(\delta)}_{\ \mathrm{loc}}}
	=
	-\qsource(x)
	\partial_\beta
	\fder{}{\qsource(x)}
	\eval{\qsource \cdot \quasi^{(\delta)}_{\ \mathrm{loc}}
	\cdots \qsource \cdot \quasi^{(\delta)}_{\ \mathrm{loc}}}
\ee
where, in accord with the notation of~\cite{HO-remarks} (which we largely follow throughout this paper)
\[
	\qsource \cdot \quasi \equiv \Int{x} \qsource(x) \quasi(x)
.
\]
Next, we sum over $n$ and observe that the result can be cast in the form
\be
	\partial_\alpha \mathcal{T}^{(\mathrm{Sch})}_{\alpha \beta} = - \qsource \; \partial_\beta \fder{\dual[\qsource]}{\qsource}
,
\label{eq:EMT-intro}
\ee
where we have chosen to define $\mathcal{T}^{(\mathrm{Sch})}_{\alpha \beta}$ such that it is only contains connected contributions (multiplying by $e^{\dual[\qsource]}$ restores the disconnected pieces). A crucial point is that (modulo some subtleties to be dealt with later) we interpret $\mathcal{T}^{(\mathrm{Sch})}_{\alpha \beta}$ as a representation of the energy-momentum tensor; the `Sch' serves to reminds us that this representation involves the Schwinger functional. When working in an arbitrary representation, we will utilize the symbol $\mathcal{T}_{\alpha \beta}$.

Evidence will be assembled for this in \sect{EM} as follows. As we will see in \sect{proposal}, the Schwinger functional representation makes it particularly transparent that the two main ingredients on the \rhs\ of~\eq{EMT-intro}, $\qsource$ and $\delta \dual/ \delta \qsource$ are (again, modulo some subtleties to be discussed later) representations of a pair of conformal primary fields with scaling dimensions, $\D-\delta$ and $\delta$:
\begin{align}
	\quasi_{\qsource}^{(\D-\delta)}(x) & = \qsource(x)
,
\\
	\quasi_{\qsource}^{(\delta)}(x) & = \fder{\dual[\qsource]}{\qsource(x)}
.
\end{align}
The subscript $\qsource$ adorning the conformal primary fields is a reminder that we are considering a particular representation. Let us emphasise again that this particular representation of $\quasi^{(\delta)}$ is non-local and that we will see later how to obtain a reassuringly local representation, via the ERG.

Existence of $\mathcal{T}_{\alpha \beta}$ is established in \sect{Existence}. The basic idea is to exploit the fact that $\quasi_{\qsource}^{(\D-\delta)}$ and $\quasi_{\qsource}^{(\delta)}$ can be combined in various ways to express translation, rotation and dilatation invariance. For example, translational invariance of the Schwinger functional can be stated as
 \[
 	\partial_\beta \qsource \cdot \fder{\dual[\qsource]}{\qsource} = 0,
 \]
 which suggests the existence of a tensor field such that
 \[
 	\quasi_{\qsource}^{(\D-\delta)}(x) \partial_\beta \quasi_{\qsource}^{(\delta)}(x)
	=
	-\partial_\alpha F_{\alpha \beta}(x)
.
 \]
(As will be discussed more fully later, this is not quite the full story: existence of a quasi-local representation of the theory is required.) Rotational invariance implies symmetry of $F_{\alpha \beta}$. 
Furthermore, it will be shown, under certain conditions, that $F_{\alpha\beta}$ can be `improved' in such a way that its trace corresponds to the Ward identity associated with dilatation invariance. Assuming this improvement to be done, conformal invariance is confirmed in \sect{Conditions}.

This analysis of the improvement procedure will be seen to have close parallels to that of Polchinski's classic paper~\cite{Pol-ScaleConformal}. Indeed, as in the latter, a sufficient condition for this improvement is essentially that primary vector fields of scaling dimension $\D-1$ are absent from the spectrum. In the same paper, Polchinski completed an argument due to Zamolodchikov~\cite{Zam-cfn} showing that, in two dimensions, the energy-momentum tensor of a scale-invariant theory can be rendered traceless if the theory is unitary and the spectrum of fields is discrete. The veracity of this for $\D>2$ has been much debated~\cite{Nakayama-ScaleConformal}. 

As additional confirmation of the consistency of our approach it will be shown 
in \sect{ConformalCovariance}  that $\mathcal{T}_{\alpha \beta}$ has the same properties under conformal transformations as a tensor, conformal primary field of scaling dimension, $\D$. (We use the term field with care since the associated representation is non-local, but henceforth will be less assiduous.) Let us emphasise that we are not claiming $\eval{T_{\alpha\beta}}$ is a representation of the energy-momentum tensor in the sense of having the correct transformation properties under the appropriate representation of the conformal group; only for the full $\mathcal{T}_{\alpha \beta}$ does this hold.

While this section began with the standard representation of the energy-momentum tensor---i.e.\ a quasi-local object---from the perspective of this paper we view as more primitive the non-local representation furnished by the Schwinger functional, $\mathcal{T}_{\alpha \beta}$. For theories supporting a quasi-local representation, $T_{\alpha\beta}$ can be recovered via the ERG, as will become apparent in \sect{qlrep}. With this achieved, another of the central results of this paper will become apparent: in the ERG representation, the trace of the energy-momentum tensor is nothing but the exactly marginal, redundant field possessed by every critical fixed-point. (Redundant fields correspond to quasi-local field redefinitions.)  It is the existence of this field which causes quasi-local fixed-point theories to divide up into equivalence classes: every fixed-point theory exists as a one-parameter family of physically equivalent theories~\cite{Wegner-CS,WegnerInv,TRM-Elements,Fundamentals,RGN}. This is the origin of the quantization of the spectrum of $\delta$.

The construction of the energy-momentum tensor will be illustrated in \sect{GFP} using the example of the Gaussian fixed-point; \sect{non-unitary} demonstrates how the construction breaks down for a simple, non-unitary theory.

\section{Conformal Symmetry in QFT}

\subsection{Elementary Properties of the Conformal Group}

In this section, we recall some basic features of the conformal group; henceforth, unless stated otherwise, we work in Euclidean space. The generators $\{\Trans_\mu,  \Rot_{\mu\nu}, \Dil, \SCT_\mu \}$ respectively generate translations, rotations, dilatations (scale transformations) and special conformal transformations;
the non-zero commutators are:
\begin{align}
	\bigl[\Dil,\Trans_\mu \bigr] & = \Trans_\mu,
	&
	\bigl[\Rot_{\mu\nu}, \Rot_{\sigma\rho} \bigr]
	&= 
	\delta_{\mu\sigma} \Rot_{\nu\rho} - \delta_{\nu\sigma} \Rot_{\mu\rho}
	-\delta_{\mu\rho} \Rot_{\nu\sigma} + \delta_{\nu\rho} \Rot_{\mu\sigma},
\nonumber
\\
	\bigl[\Rot_{\mu\nu}, \Trans_\sigma \bigr] &= \delta_{\mu\sigma} \Trans_\nu - \delta_{\nu\sigma} \Trans_\mu,
	&
	\bigl[\Rot_{\mu\nu}, \SCT_\sigma \bigr] & = \delta_{\mu\sigma} \SCT_{\nu} - \delta_{\nu\sigma} \SCT_{\mu},
\nonumber
\\
	\bigl[\Dil,\SCT_\mu \bigr] &=  -\SCT_\mu,
	&
	\bigl[\SCT_\mu, \Trans_\nu\bigr] &= 2\delta_{\mu\nu} \Dil + 2\Rot_{\mu\nu}
.
\label{eq:ConformalAlgebra}
\end{align}
Though it will not exploited in this paper, it is worth noting that the commutation relations can be recast in a manner which makes explicit the isomorphism between the conformal group and $SO(\D+1,1)$ (see, for example, \cite{YellowPages}).

A scalar conformal primary field, $\mathscr{O}(x)$, with scaling dimension $\Delta$, satisfies%
\footnote{Were we to work in an operator formalism, the expressions on the {\lhs}s would appear
as commutators.}
\be
	\Trans_\mu \quasi = \partial_\mu \quasi,
	\qquad
	\Rot_{\mu\nu}\quasi = \rot{\mu\nu} \quasi,
	\qquad
	\Dil \quasi= \dil{\Delta} \quasi,
	\qquad
	\SCT_\mu\quasi = \sct{\Delta}{\mu} \quasi,
\label{eq:qpo}
\ee
where $\{\partial_\mu, \rot{\mu\nu}, \dil{\Delta}, \sct{\Delta}{\mu} \}$ are taken such that
they satisfy a version of the commutation relations above in which the signs are flipped.
This is crucial if~\eq{qpo} is to be consistent with the commutation relations. For example, it follows from~\eq{qpo} that
\be
	\bigl[\Dil,\SCT_\mu \bigr]\quasi(x)
	=
	\bigl[\sct{\Delta}{\mu}, \dil{\Delta}\bigr]  \quasi(x) 
,
\ee
from which we deduce that $\bigl[\dil{\Delta}, \sct{\Delta}{\mu}\bigr] = + \sct{\Delta}{\mu}$, as
compared with $\bigl[\Dil,\SCT_\mu \bigr] =  -\SCT_\mu$ in~\eq{ConformalAlgebra}.
With this in mind, we take:
\begin{subequations}
\begin{align}
	\rot{\mu\nu} \quasi(x) & = \bigl( x_\mu \partial_\nu - x_\nu \partial_\mu \bigr)\quasi(x),
\label{eq:rot-qpo}
\\
	\dil{\Delta} \quasi(x) & = \bigl(x\cdot\partial + \Delta \bigr) \quasi(x),
\label{eq:dil-qpo}
\\
	\sct{\Delta}{\mu} \quasi(x)
	&=
	\bigl(
		2x_\mu \bigl( x\cdot \partial + \Delta \bigr)
		-x^2 \partial_\mu
	\bigr) \quasi(x)
.
\label{eq:sct-qpo}
\end{align}
\end{subequations}
The general modification of~\eq{qpo} appropriate to non-scalar fields can be found in~\cite{YellowPages}. For our purposes, we explicitly give the version appropriate to tensor fields:
\begin{subequations}
\begin{align}
	\Rot_{\mu\nu} \quasi_{\alpha_1 \ldots \alpha_n}(x) & =
	\rot{\mu\nu} \quasi_{\alpha_1 \ldots \alpha_n}(x) 
	+
	\sum_{i=1}^n 
	\bigl(
		\delta_{\mu \alpha_i} \delta_{\gamma\nu} 
		- \delta_{\nu \alpha_i} \delta_{\gamma \mu}
	\bigr)
	 \quasi_{\alpha_1 \ldots \gamma \ldots \alpha_n}(x)
	,
\label{eq:rot-qpo-tensor}
\\
	\SCT_\mu \quasi_{\alpha_1 \ldots \alpha_n}(x)
	& = \sct{\Delta}{\mu} \quasi_{\alpha_1 \ldots \alpha_n}(x)
	+2 
	\sum_{i=1}^n 
	\bigl(
		\delta_{\mu \alpha_i} \delta_{ \gamma \beta} 
		-\delta_{\beta \alpha_i } \delta_{\gamma\mu} 
	\bigr)
	x_{\beta} \quasi_{\alpha_1 \ldots \gamma\ldots \alpha_n}(x)
.
\label{eq:sct-qpo-tensor}
\end{align}
\end{subequations}
These relationships will play an important role when we deal with the energy-momentum tensor in \sect{EM}.

Let us emphasise that, at this stage, the representation of the $\{\Trans_\mu,  \Rot_{\mu\nu}, \Dil, \SCT_\mu \}$ and the $\mathscr{O}(x)$ are yet to be fixed; a key theme of this paper will be the exploration of certain representations thereof, some of which are non-standard. 

\subsection{Correlation Functions}
\label{sec:correlation}

In the context of QFT, the chief consequence of
conformal symmetry is that various correlation functions are annihilated by $\{\partial_\mu, \rot{\mu\nu}, \dil{\Delta}, \sct{\Delta}{\mu} \}$.
Specifically, correlation functions involving only the conformal primaries are annihilated by all members of the set, whereas those involving descendant fields (the derivatives of the conformal primaries) are annihilated only by those corresponding to translations, rotations and dilatations.
Thus we have, for all $n$,
\be
	\biggl(\sum_{j=1}^n \sct{\Delta_{i_j}}{\mu} (x_j)\biggr)
	\ \eval{\quasi_{i_1}(x_1) \cdots \quasi_{i_n}(x_n)} = 0
,
\qquad
	\forall\ i_1, \ldots, i_n
\label{eq:sct-cov}
\ee
whereas the remaining conditions read%
\footnote{%
Indices near the beginning of the alphabet are understood to label all fields, rather than just the conformal primaries.
}
, now for all $a_1, \ldots, a_n$:
\be
\begin{split}
	\biggl(\sum_{j=1}^n \pder{}{x_{j\mu}}\biggr) \
	\eval{\operator_{a_1}(x_1) \cdots \operator_{a_n}(x_n)} = 0,
\\
	\biggl(\sum_{j=1}^n \rot{\mu\nu}(x_j)\biggr) \
	\eval{\operator_{a_1}(x_1) \cdots \operator_{a_n}(x_n)} = 0,
\\
	\biggl(\sum_{j=1}^n \dil{\Delta_{i_j}}(x)\biggr) \
	\eval{\operator_{a_1}(x_1) \cdots \operator_{a_n}(x_n)} = 0
.
\label{eq:rest-cov}
\end{split}
\ee
The ultimate aim is to find solutions to~\eqs{sct-cov}{rest-cov} that correspond to acceptable QFTs. 

A key step for what follows is to introduce a set of sources, $\qsource_i(x)$, conjugate to the conformal primary fields $\quasi_i(x)$ (any Euclidean indices are suppressed). There is no need to introduce sources for the descendants since the associated correlation functions can be generated from the analogue involving just primaries by acting with appropriate derivatives.
Restricting to conformal primary fields, we tentatively rewrite~\eqs{sct-cov}{rest-cov} as
\begin{subequations}
\begin{align}
	\biggl(\sum_i \sct{\D-\scaling_i}{\mu} \ \qsource_i \cdot \fder{}{\qsource_i} 
	\biggr)
	e^{\dual[\{\qsource\}]} & = 0,
\label{eq:annihilate-SCT}
\\
	\biggl(\sum_i \dil{\D-\scaling_i}\qsource_i \cdot \fder{}{\qsource_i} 
	\biggr)
	e^{\dual[\{\qsource\}]} & = 0,
\label{eq:annihilate-DIL}
\\
	\biggl(\sum_i \partial_\mu\qsource_i \cdot \fder{}{\qsource_i} 
	\biggr)
	e^{\dual[\{\qsource\}]} & = 0,
\label{eq:annihilate-TRANS}
\\
	\biggl(\sum_i \rot{\mu\nu}\qsource_i \cdot \fder{}{\qsource_i} 
	\biggr)
	e^{\dual[\{\qsource\}]} & = 0,
\label{eq:annihilate-ROT}
\end{align}
\end{subequations}
where
\be
	e^{\dual[\{\qsource\}]} = \eval{e^{\sum_i \qsource_i \cdot \quasi_i}} 
\ee

In general, considerable care must be taken defining the expectation value of exponentials, due to both IR and UV singularities. However, this paper will only directly utilize expectation values involving $\qsource$ (which, we recall, can loosely be thought of as coupling to the lowest dimension conformal primary field); indeed, for brevity we will henceforth deal only with this single source, the scaling dimension of which is $\D-\delta$ (it is a simple matter to insert the remaining sources, should one so desire).

At certain stages, we will simply assume that the Schwinger functional involving solely $\qsource$, $\dual[\qsource]$, is well defined. To be precise, when we talk of existence of the Schwinger functional, it is meant that the correlation functions of the field conjugate to $\qsource$ can be directly subsumed into  $\dual[\qsource]$ and so the \naive\ identities~\eq{annihilate-SCT}--\eq{annihilate-ROT} hold. Note that existence of $\dual[\qsource]$ is considered a separate property from $\dual[\qsource]$ being non-zero.

By definition, we take $\qsource_i\cdot \quasi_i $ to have  zero scaling dimension; this implies that $\qsource_i(x)$ has scaling dimension $\D-\scaling_i$.
This leads us to the first of several functional representations of the conformal generators that will be presented in this paper.
\begin{representation} Schwinger Functional Representation
\label{rep:J}
 \begin{subequations}
\begin{align}
	\Trans_\mu & = \partial_\mu \qsource \cdot \fder{}{\qsource} ,
	& \Rot_{\mu\nu} & =  L_{\mu\nu}\qsource \cdot \fder{}{\qsource}, 
\label{eq:TransRot-Source}
\\
	\Dil & = \dil{\D-\delta} \qsource \cdot \fder{}{\qsource},
	&
	\SCT_\mu & = \sct{\D-\delta}{\mu}  \qsource \cdot \fder{}{\qsource}.
\label{eq:DilSCT-Source}
\end{align}
\end{subequations}
It is easy to check that these generators satisfy the conformal algebra by utilizing~\eq{rot-qpo}, \eqs{dil-qpo}{sct-qpo}, together with the following relationships which follow from integrating by parts:
 \begin{subequations}
\begin{align}
	\partial_\mu \qsource \cdot \fder{}{\qsource} & 
	= - \qsource \cdot \partial_\mu \fder{}{\qsource},
	\qquad
	&L_{\mu\nu} \qsource \cdot \fder{}{\qsource} & 
	= -  \qsource \cdot L_{\mu\nu} \fder{}{\qsource},
\\
	 \dil{\D-\delta} \qsource \cdot \fder{}{\qsource} &
	 = -\qsource\cdot \dil{\delta} \fder{}{\qsource},
	 \qquad
	 &\sct{\D-\delta}{\mu}  \qsource \cdot \fder{}{\qsource} &
	 = -  \qsource \cdot \sct{\delta}{\mu}\fder{}{\qsource}
.
\end{align}
 \end{subequations}
Note that the fact that $\{\partial_\mu, \rot{\mu\nu}, \dil{\Delta}, \sct{\Delta}{\mu} \}$ satisfy a version of the conformal commutation relations in which the order of the commutators is flipped is crucial.
\end{representation}

Now that we have a concrete functional representation of the conformal algebra, it is appropriate to mention a subtlety pertaining to volume terms. To illustrate this issue, consider the effect of the dilatation operator on an integrated field. Recalling~\eqs{qpo}{dil-qpo}, it is apparent that we expect
\be
	\Dil \Int{x} \quasi(x) = (\Delta -\D) \Int{x} \quasi(x)
\label{eq:integrated}
.
\ee
In deriving this, we have implicitly assumed that $\quasi$ depends on a field which dies of sufficiently rapidly at infinity. However, for the identity operator this is not the case. To match the two sides of~\eq{integrated} in this situation---and bearing in mind that $\Delta=0$---suggests that the dilatation generator should be supplemented by a term
\[
	-\D V \pder{}{V},
\]
with $V$ being the volume of the space on which the field theory lives. For this paper, however, we will generally ignore volume terms; as such, we henceforth understand equality in functional equations to hold only up to volume terms. This issue will be addressed more fully in~\cite{OJR-Volume}.

\subsection{From Sources to the Fundamental Field}
\label{sec:SourcesToFields}

Up until this point, our functional representation has utilized sources. The transition to fields proceeds in several steps, along the way giving new representations of the conformal algebra. The first such step is provided by the shift~\eq{J-shift}. Clearly, at this stage, the dependence on $\qsource$ and $\dfield$ will not be independent. However, the link will be severed in a subsequent representation. To prepare for this severing, our aim in this section is, given the shift~\eq{J-shift}, to construct a representation in which the generators involve functional derivatives \wrt\ $\dfield$ (rather than $\partial^2 \dfield$).

As mentioned earlier, we anticipate that $\dfield$ will play the role of the fundamental field. Before proceeding, it is worth pointing out that there is a subtlety over precisely what is meant by the latter. Strictly speaking, both the Wilsonian effective action and the field to which $\qsource$ couples are \emph{built} out of $\dfield$. Within the ERG representation (and assuming sufficiently good IR behaviour), $\dfield$ coincides with a conformal primary field only up to non-universal terms, which vanish in the limit that the regularization is removed. While this subtlety will be largely glossed over since it seems to have no great significance, the issue of theories for which bad IR behaviour prevents $\dfield$ from corresponding to a conformal primary in any sense will be returned to, later.

With the aim of producing a representation of the generators  involving functional derivatives \wrt\ $\dfield$, we exploit the commutators
\be
\begin{split}
	\bigl[\partial^2, \dil{\Delta}\bigr]
	& = 2 \partial^2
,
\\
	\bigl[\partial^2, \sct{\Delta}{\mu}\bigr]
	& = 4(\Delta - \delta_0) \partial_\mu + 4 x_\mu \partial^2
,
\end{split}
\label{eq:d^2-comms}
\ee
where $\delta_0$ (which we recognize as the canonical dimension of the fundamental field) is given by
\be
	\delta_0 \equiv \frac{\D - 2}{2}
.
\label{eq:canonical}
\ee
Next, define $\ep_0$ to be Green's function for $-\partial^2$:
\be
	-\partial^2 \ep_0(x) = \DD{\D}(x)
.
\label{eq:GreenFunction}
\ee
Employing notation such that, for fields $\dfield(x)$, $\psi(x)$ and kernel $\cutoff(x,y)$
\be
	\dfield \cdot \cutoff \cdot \psi \equiv \Int{x} \volume{y} \dfield(x) \cutoff(x,y) \psi(y)
,
\ee
observe that
\begin{align}
	\dil{\D-\delta} \qsource \cdot \fder{}{\qsource} \dual[\qsource] 
	\Biggr\vert_{\qsource=\partial^2 \dfield}
	&
	=
	\dil{\D-\delta} (\partial^2 \dfield) \cdot \fder{}{(\partial^2 \dfield)} \dual[\partial^2 \dfield] 
\nonumber
\\
	&
	=
	-\dil{\D-\delta} (\partial^2 \dfield) \cdot \ep_0 \cdot \fder{}{\dfield} \dual[\partial^2 \dfield] 
\nonumber
\\
	&
	=
	\Bigl(
		\bigl[\partial^2, \dil{\D-\delta}\bigr]\dfield -\partial^2 \bigl( \dil{\D-\delta} \dfield)
	\Bigr)
	 \cdot \ep_0 \cdot \fder{}{\dfield} \dual[\partial^2 \dfield] 
\nonumber
\\
	&
	=
	\dil{\D-\delta-2}
	\dfield \cdot \fder{}{\dfield} \dual[\partial^2 \dfield] 
,
\end{align}
where we recall from the introduction that $\delta$ is the scaling dimension of the fundamental field. 
Performing similar manipulations for the special conformal generator we arrive at:
\begin{subequations}
\begin{align}
	\dil{\D-\delta} \qsource \cdot \fder{}{\qsource} \dual[\qsource] 
	\Biggr\vert_{\qsource=\partial^2 \dfield}
	&
	=
	\dil{\delta - \eta} \dfield \cdot \fder{}{\dfield} \dual[\partial^2 \dfield]
,
\\
	\sct{\D-\delta}{\mu} 
	\qsource \cdot \fder{}{\qsource}  \dual[\qsource] \Biggr\vert_{\qsource=\partial^2 \dfield}
	&
	=
	\sct{\delta - \eta}{\mu} \dfield \cdot \fder{}{\dfield}  \dual[\partial^2 \dfield]
	-
	2 \eta\,
	\partial_\mu \dfield \cdot \ep_0 \cdot \fder{}{\dfield}
	\dual[\partial^2 \dfield]
,
\label{eq:ParaPrelim-SCT}
\end{align}
\end{subequations}
where we have introduced the anomalous dimension, $\eta$, defined via
\be
	\delta = \delta_0 + \eta /2
.
\label{eq:delta_star}
\ee

\begin{representation} Para-Schwinger Functional Representation
\begin{subequations}
\begin{alignat}{2}
	\Trans_{\mu} & = \TransR_\mu && \equiv
	\partial_\mu \qsource \cdot \fder{}{ \qsource }
	+
	\partial_\mu \dfield \cdot \fder{}{\dfield}
,
\label{eq:paraP}
\\
	\Rot_{\mu\nu} &= \RotR_{\mu\nu} && \equiv
	L_{\mu\nu}  \qsource \cdot \fder{}{ \qsource}
	+
	\rot{\mu\nu} \dfield \cdot \fder{}{\dfield}
,
\label{eq:paraM}
\\
	\Dil & = \DilR && \equiv
	\dil{\D-\delta}  \qsource \cdot \fder{}{ \qsource}
	+
	\dil{\delta - \eta} \dfield \cdot \fder{}{\dfield}
,
\label{eq:paraD}
\\
	\SCT_{\mu} 
	&
	= \SCTR_\mu && \equiv
	\sct{\D-\delta}{\mu}   \qsource  \cdot \fder{}{ \qsource }
	+
	\sct{\delta - \eta}{\mu} \dfield \cdot \fder{}{\dfield}  
		-2\eta \, \partial_\mu \dfield \cdot \ep_0 \cdot \fder{}{\dfield},
\label{eq:paraK}
\end{alignat}
\end{subequations}
where $\TransR_\mu$, $\RotR_{\mu\nu},\ldots$ correspond to the expressions for the various generators in the present representation. It is straightforward to confirm from~\eq{rot-qpo}, \eqs{dil-qpo}{sct-qpo}, together with translational and rotational invariance of $\ep_0$, that these generators satisfy the requisite commutation relations. In this representation,  a conformal field theory is such that each of these generators annihilates $\dual[\qsource + \partial^2 \dfield]$.
\label{rep:para}
\end{representation}

Before introducing the next representation, it is worth mentioning that the functional $\dual[\qsource + \partial^2 \dfield]$ may have different (quasi)-locality properties with respect to $\qsource$ and $\dfield$. For non-trivial fixed-points this will not be the case, as can be seen at the two-point level. In momentum space, the two-point correlation function goes like $1/p^{2(1-\eta/2)}$. For non-trivial fixed-points, $\eta/2$ is some non-integer number. While multiplying by a factor of $p^2$ removes the divergence as $p^2 \rightarrow 0$, it does not remove the non-locality. For trivial fixed-points, however, non-locality may be ameliorated. This can be convenient and is exploited in \sect{GFP}.

\subsection{From The Fundamental Field to the ERG}
\label{sec:FieldsToERG}

The aim now is to go from this representation to one in which the dynamics of the theory is encoded in some auxiliary object. To begin, we introduce an auxiliary functional, $\mathcal{U}$, which, save for insisting translational and rotational invariance, we leave arbitrary for now. From here, we define:
\be
	\dual_\mathcal{U}[\dfield, \qsource] \equiv -\dual[\qsource + \partial^2\dfield] + \mathcal{U}[\dfield, \qsource]
.
\label{eq:W_U[j]}
\ee
Notice that $\dual_\mathcal{U}[\dfield, \qsource]$ may depend independently on $\qsource$ and $\dfield$.
The arbitrariness in $\mathcal{U}$ is a manifestation of the freedom inherent in constructing ERGs, which has been recognized since the birth of the subject~\cite{Wegner-CS,WegnerInv}. As particularly emphasised in~\cite{TRM+JL}, this can be understood from deriving the ERG equation via a quasi-local field redefinition under the path integral, which will be elaborated upon at the end of this section. In \sect{Pol} we will focus on a particular choice of $\mathcal{U}$ which turns out to reproduce what is essentially Polchinski's ERG equation. 

The idea now is to encode the dynamics in an object, $\Sintd[\dfield, \qsource]$, and introduce an operator, $\op$ (about which more will be said, below), such that
\be
	e^{\op} e^{-\Sintd[\dfield, \qsource]} = e^{-\dual_\mathcal{U}[\dfield, \qsource]}
.
\label{eq:dualFromS}
\ee
Thus, given $\Sintd$ we can, in principle, recover the correlation functions.
In this sense, $\Sintd$ encodes the dynamics of the theory. 
It should be pointed out that any vacuum contributions to the Wilsonian effective action are unconstrained within our approach. The conditions on $\dual_\mathcal{U}$ implied by conformal invariance are blind to vacuum terms. Consequently, we are free to add any vacuum term we like to $\mathcal{U}$ which amounts, via~\eq{dualFromS}, to an arbitrary vacuum contribution to the Wilsonian effective action.

Before moving on let us not that, in general, the Wilsonian approach deals not just with scale (or conformally) invariant theories but with theories exhibiting scale dependence. Scale-independent actions are typically denoted by $\Sintd_\star$ and solve the fixed-point version of an ERG equation. However, since this paper will only ever deal with fixed-point quantities, the $\star$ will henceforth be dropped.

There is an implicit assumption that it is possible to find a non-trivial $\op$ such that~\eq{dualFromS} exists. Given this, 
a representation can be constructed as follows. Given a generator, $g$, and a representation of this generator, denoted by $\mathscr{G}$, define
\be
	\mathscr{G}_{\mathcal{U}}
	\equiv
	e^{-\op} e^{-\mathcal{U}} \overrightarrow{\mathscr{G}} e^{\mathcal{U}} e^{\op}
,
\label{eq:G_U}
\ee
where the arrow indicates that the generator acts on everything to their right, with it being understood that further terms may follow the $e^{\op}$ (without the arrow, we would take the generator just to act on the explicitly written terms to its right). By construction, if generators $\mathscr{G}$ and $\mathscr{G}'$ satisfy some commutation relation, then the same is true of $\mathscr{G}_{\mathcal{U}}$ and $\mathscr{G}'_{\mathcal{U}}$. Immediately, this allows us to construct a representation as follows.
\begin{representation} Auxiliary functional representation
\begin{subequations}
\begin{align}
	\Trans_{\mu} & = {\TransR}_{U \mu} 
,
\label{eq:auxP}
\\
	\Rot_{\mu\nu} &= {\RotR}_{U \mu\nu}
,
\label{eq:auxM}
\\
	\Dil &= \DilR_{\mathcal{U}}
,
\label{eq:auxD}
\\
	\SCT_{\mu} 
	&
	=
	\SCTR_{\mathcal{U}\mu}
.
\label{eq:auxK}
\end{align}
\end{subequations}
For a conformal field theory, in this representation, each generator annihilates $e^{-\Sintd[\dfield, \qsource]}$, as follows from \eqss{W_U[j]}{dualFromS}{G_U}.
\label{rep:aux}
\end{representation}

Let us now explore some possibilities for $\op$.
Translation invariance of the correlation functions and of $\mathcal{U}$ implies that
\be
	\biggl(
		\partial_\mu \qsource \cdot \fder{}{\qsource} + \partial_\mu \dfield \cdot \fder{}{\dfield}
	\biggr)
	e^{-\dual_\mathcal{U}[\dfield, \qsource]}
	=
	0
,
\ee
with a similar expression implied by rotational invariance. Next consider substituting for $\dual_\mathcal{U}$ using~\eq{dualFromS} and commuting $e^{\op}$ to the \lhs.
Now demand manifest translation invariance of $\Sintd$, by which we mean\footnote{It is tempting to speculate that relaxing this requirement may be illuminating in the context of lattice theories.} 
\be
	\biggl(
		\partial_\mu \qsource\cdot \fder{}{\qsource} + \partial_\mu \dfield \cdot \fder{}{\dfield}
	\biggr)
	e^{-\Sintd[\dfield, \qsource]} = 0
.
\ee
This, together with the similar constraint coming from demanding manifest rotational invariance, implies
\be
	\biggl[
		e^{\op}, \partial _\mu \dfield \cdot \fder{}{\dfield}
	\biggr] e^{\Sintd[\dfield, \qsource]}
	=
	0
,
	\qquad
	\biggl[
		e^{\op}, \rot{\mu\nu} \dfield \cdot \fder{}{\dfield}
	\biggr] e^{\Sintd[\dfield, \qsource]}
	=
	0
.
\label{eq:eY-comm}
\ee

The most obvious solution to these constraints is $\op = b \dfield \cdot \delta / \delta \dfield$, for some constant $b$. However, in this case $\Sintd[\dfield]$ is related to $\dual_\mathcal{U}$ by rescaling each leg of every vertex of the latter  by a factor of $e^{-b}$, which gives us nothing new. Instead, we solve the constraints by introducing a kernel, $\ep\bigl((x-y)^2\bigr)$, and taking
\be
	\op
	=
	\frac{1}{2}
	\fder{}{\dfield} \cdot \ep \cdot \fder{}{\dfield}
.
\label{eq:op}
\ee
Typically, $\ep$ has, roughly speaking, the form of a regularized propagator (care must be taken with this identification, as discussed in~\cite{Fundamentals}). Given a momentum space UV cutoff function, $\cutoff(p^2)$, and using the same symbol for position-space space objects and their Fourier transforms, we write the Fourier transform of $\ep$ as
\be
	\ep(p^2) = \frac{\cutoff(p^2)}{p^2}
.
\label{eq:EffProp}
\ee
Of course, we have been guided to~\eqs{op}{EffProp} by our pre-existing knowledge of both the form of the ERG equation and the (related) role of the propagator in the standard path integral approach to QFT. Let us stress that we have not derived these equations and the uniqueness or otherwise of this particular solution is an important question to answer, but beyond the scope of this paper. Given our prior knowledge of what to look for, we anticipate that~\eqs{op}{EffProp} will lead to a useful representation of the conformal algebra (as discussed in the introduction, by `useful' we mean that the constraint which picks out physically acceptable theories is easy to implement; for the ERG this constraint is quasi-locality).
Before moving on, let us mention that~\eq{EffProp} suffers from IR problems in $\D=2$.%
\footnote{%
Indeed the true propagator at the Gaussian fixed-point in $\D=2$ has logarithmic behaviour, emphasising that the interpretation of $\ep$ as a regularized propagator must be taken with a pinch of salt.
}
Strictly speaking, this suggests that in $\D=2$ we should work in finite volume, at least at intermediate stages. 
 
Observe that it is possible to simplify the expressions for $\TransR_\mu$ and $\RotR_{\mu\nu}$. Since $\mathcal{U}$ is taken to be invariant under translations and rotations, we can write
\be
	\TransR_{\mathcal{U}\mu} = e^{-\op} \overrightarrow{\TransR}_\mu e^{\op},
	\qquad
	\RotR_{\mathcal{U}\mu \nu} = e^{-\op} \overrightarrow{\RotR}_{\mu \nu} e^{\op}
.
\label{eq:similarity}
\ee
Given~\eq{eY-comm}, it is tempting to try to simplify these expressions further, but a little care must be taken. If these generators act on something which is transitionally and rotationally invariant, then $\Trans_\mu$ and $\Rot_{\mu\nu}$ are transparent to $e^{\op}$ which can be trivially commuted to the left, whereupon it is annihilated by $e^{-\op}$, leaving behind just $\Trans_\mu$. But suppose, for example, that $\TransR_\mu$ acts on something not translationally invariant, such as
\[
	A_\mu[\dfield] \equiv \hf \Int{x} \!\! \Int{y} \dfield(x) \dfield(y) (x+y)_\mu F\bigl( (x-y)^2 \bigr)
.
\] 
Using~\eq{op}, it is easy to check that
\[
	\biggl[
		\op, \partial _\mu \dfield \cdot \fder{}{\dfield}
	\biggr]
	A_\nu [\dfield]
	=\delta_{\mu\nu} \tr \ep \cdot F
.
\]
The origin of this remainder relates to the discussion under~\eq{integrated}; indeed, for many purposes of interest it is consistent to take $\TransR_{\mathcal{U}\mu}  = \TransR_{\mu}$.

Finally,  we construct a representation in which the generators of dilatations and special conformal transformations contain the action. Recalling~\eq{G_U}, let us define
\be
	\mathscr{G}_{\Sintd}
	\equiv
	e^{\Sintd} \overrightarrow{\mathscr{G}}_{\mathcal{U}} e^{-\Sintd}
.
\label{eq:gen_S}
\ee
By construction it is apparent that, if $\mathscr{G}_{\mathcal{U}}$ and $\mathscr{G}'_{\mathcal{U}}$ satisfy some commutation relation, then so too do $\mathscr{G}_{\Sintd}$ and $\mathscr{G}'_{\Sintd}$, leading to the next representation.
 \begin{representation} ERG representation
\begin{subequations}
\begin{align}
	\Trans_{\mu} & = {\TransR}_{\Sintd \mu} 
,
\label{eq:actionP}
\\
	\Rot_{\mu\nu} &= {\RotR}_{\Sintd \mu\nu}
,
\label{eq:actionM}
\\
	\Dil &= \DilR_{\Sintd}
,
\label{eq:actionD}
\\
	\SCT_{\mu} 
	&
	=
	\SCTR_{\Sintd \mu}
.
\label{eq:actionK}
\end{align}
\end{subequations}

\label{rep:ERG}
\end{representation}

We will now study some of the properties of the last two representations. As remarked above, in the auxiliary functional representation, a conformal field theory is such that $e^{-\Sintd}$ is annihilated by each of the generators. For translations and rotations, this implies that $\Sintd$, itself, is thus annihilated. However, the same is not true for dilatations and special conformal transformations. In the ERG representation, the associated constraints are most naturally expressed as
\begin{subequations}
\begin{align}
\label{eq:dilConstraint}
	E_{\Sintd}[\dfield, \qsource] 
	& = e^{\Sintd[\dfield, \qsource]} \DilR_{\mathcal{U}} e^{-\Sintd[\dfield, \qsource]} = 0,
\\
\label{eq:sctConstraint}
	E_{\Sintd \mu }[\dfield, \qsource] 
	& = e^{\Sintd[\dfield, \qsource]}\SCTR_{\mathcal{U} \mu} e^{-\Sintd[\dfield, \qsource]} = 0
.
\end{align}
\end{subequations}
These translate into non-linear constraints on $\Sintd$, as we will see in an explicit example in the next section. 

Indeed, Given certain restrictions (pertaining to quasi-locality) to be discussed in \sect{Pol}, \eq{dilConstraint} will be recognized as nothing but an ERG equation (in the presence of sources) specialized to a fixed-point. Equation~\eq{sctConstraint} is an additional constraint on the action enforcing conformal invariance, along the lines of~\cite{Schafer-Conformal}. If we choose to restrict $\mathcal{U}$ to depend only on $\dfield$, then the only source dependence in~\eq{dilConstraint} occurs through the action and the $\quasi_i$ can be picked out of the latter in a simple manner. Anticipating this, let us reinstate all sources and write
\be
	\Sintd[\dfield, \{\qsource\}] = \Sintd[\dfield] - \sum_i \qsource_i \cdot \quasi_i + \ldots
\label{eq:LinearInJ}
\ee
Substituting into~\eqs{dilConstraint}{sctConstraint}, it is apparent that
\begin{subequations}
\begin{align}
	\DilR_{\Sintd}[\dfield] \quasi_i(x) & = \dil{\Delta_i}  \quasi_i(x),
\label{eq:dilOp}
\\
	\SCTR_{\Sintd\mu}[\dfield]\quasi_i(x)& = \sct{\Delta_i}{\mu}  \quasi_i(x),
\label{eq:sctQuasi}
\end{align}
\end{subequations}
where $\DilR_{\Sintd}[\dfield]$ and $\SCTR_{\Sintd\mu}[\dfield]$ correspond to the pieces of $\DilR_{\Sintd}$ and $\SCTR_{\Sintd\mu}$ which remain when the source is set to zero.
This pair of equations confirms our expectation that the sources are conjugate to the fields, as expected. Note that the constraint of quasi-locality is necessary for the promotion of these equations to eigenvalue equations for the scaling dimensions, $\Delta_i$, as mentioned in \sect{ERG-intro}.

Let us conclude this section by discussing in a little more detail how the freedom inherent in $\mathcal{U}$ is related to the freedom inherent in the ERG.
For~\eq{dilConstraint} to correspond to a \emph{bona-fide} ERG equation, $\DilR_U$ must on the one hand be quasi-local and, on the other, must be such that (up to vacuum terms), \eq{dilConstraint} can be cast in the form~\cite{TRM+JL}
\be
	\fder{}{\dfield} \cdot \bigl( \Psi e^{-\Stotd[\dfield, J]} \bigr) = 0
,
\ee
where
\be
	\Stotd[\dfield] \equiv \hf \dfield \cdot \ep^{-1} \cdot \dfield + \Sintd[\dfield]
\label{eq:TotalAction}
\ee
and $\Psi$ (which itself depends on the action) is quasi-local.  It will be apparent from the next section that the constraint of quasi-locality rules out the apparently simplest choice $\mathcal{U}=0$.

\section{Polchinski's Equation from the Conformal Algebra}
\label{sec:Pol}

Our treatment so far has been very general; in this section we will provide a concrete realization of our ideas by showing how to derive what is essentially Polchinski's equation~\cite{Pol}. 
Mimicking the previous section, we will take the Auxiliary Functional Representation as our starting point, deriving the generators in this representation. The constraint equation~\eq{dilConstraint}
will then be seen to produce the desired ERG equation, with~\eq{sctConstraint} producing its special conformal partner. Finally, we will give the expressions for the associated conformal generators corresponding to the ERG representation.

To this end, we take $\mathcal{U}$ to be bi-linear in the field and, for brevity, work with a single source:
\be
	\dual_h[\dfield, \qsource] = -\dual[ \qsource+\partial^2 \dfield] + \hf \dfield \cdot h \cdot \dfield,
\label{eq:W_h}
\ee
where $h$ will be specified momentarily. Recalling~\eqss{paraD}{G_U}{auxD}, it is apparent that
\be
	\DilR_h =
	\dil{\D-\delta} \qsource \cdot \fder{}{\qsource}
	+
	e^{-\op} e^{-\hf \dfield \cdot h \cdot \dfield} 
	\biggl(
		\dil{\delta-\eta} \dfield \cdot \fder{}{\dfield}
	\biggr)
	e^{\hf \dfield \cdot h \cdot \dfield}  e^{\op}
,
\ee
where $\DilR_h$ stands for $\DilR_{\mathcal{U}}$, given the special choice~\eq{W_h}. To process this, let us begin by noting that
\be
	\biggl[
		\dil{\delta-\eta} \dfield \cdot \fder{}{\dfield}
	,
		e^{\hf \dfield \cdot h \cdot \dfield}
	\biggr]
	=
	-e^{\hf \dfield \cdot h \cdot \dfield}
	\hf\dfield
	\cdot
		\bigl(
			\dil{\D-\delta+\eta} h + h \dilL{\D-\delta+\eta}
		\bigr)
	\cdot
	\dfield
,
\ee
where we understand (as in~\cite{HO-remarks}) $\dil{\Delta} h + h \dilL{\Delta}$ to be shorthand for
\[
	\bigl(
		x\cdot \partial_x + y \cdot \partial_y + 2 \Delta
	\bigr)
	h(x,y)
.
\]
Choosing $h$ such that
\be
	\dil{\D-\delta+\eta} h + h \dilL{\D-\delta+\eta}
	=
	\eta \, \ep^{-1}
,
\label{eq:heq}
\ee
the expression for $\DilR_h$ becomes:
\be
	\DilR_h
	=
	\dil{\D-\delta} \qsource \cdot \fder{}{\qsource}
	+
	e^{-\op}
	\biggl(
		\dil{\delta-\eta} \dfield \cdot \fder{}{\dfield}
		-\frac{\eta}{2}
		\dfield
		\cdot
		\ep^{-1}
		\cdot
		\dfield
	\biggr)
	e^{\op}
.
\label{eq:Dil_h-prelim}
\ee

Next, define $G$ according to
\be
	\bigl(
		\D + x\cdot \partial_x + y \cdot \partial_y
	\bigr)
	\cutoff \bigl((x-y)^2\bigr)
	= \partial^2_x G\bigl((x-y)^2\bigr)
\label{eq:G}
\ee
which, in momentum space, translates to $G(p^2) = 2 \,d K(p^2)/dp^2$. This implies
\be
	G = -\bigl(\dil{\delta_0} \ep + \ep \dilL{\delta_0} \bigr)
\label{eq:Gb}
\ee
from which we observe that
\be
	\biggl[
		e^{-\op}
		,
		\dil{\delta - \eta} \dfield \cdot \fder{}{\dfield}
		-\frac{\eta}{2} \dfield \cdot \ep^{-1} \cdot \dfield
	\biggr]
\\
	=
	\biggl(
	\hf 
	\fder{}{\dfield}  
	\cdot 
	G
	\cdot \fder{}{\dfield}
	+ \eta\, \dfield \cdot \fder{}{\dfield}
	\biggr)
	e^{-\op}
\label{eq:commutator-dil}
\ee
where, following the discussion under~\eq{integrated}, a volume term has been discarded. Substituting back into~\eq{Dil_h-prelim} yields the final expression for the dilatation generator in this representation:
\be
	\DilR_h
	=
	\dil{\D-\delta} \qsource \cdot \fder{}{\qsource}
	+
	\dil{\delta} \dfield \cdot \fder{}{\dfield}
	+
	\hf 
	\fder{}{\dfield}  
	\cdot 
	G
	\cdot \fder{}{\dfield}
	-\frac{\eta}{2} \dfield \cdot \ep^{-1} \cdot \dfield
.
\label{eq:D_h}
\ee
The constraint equation~\eq{dilConstraint} now yields the 'canonical' ERG equation, specialized to a fixed-point:
\be
	\biggl(
	\dil{\D-\delta} \qsource \cdot \fder{}{\qsource}
	+
	\dil{\delta} \dfield \cdot \fder{}{\dfield}
	\biggr)
	\Sintd[\dfield, \qsource]
	=
	\hf
	\fder{\Sintd}{\dfield} \cdot G \cdot \fder{\Sintd}{\dfield}
	-
	\hf \fder{}{\dfield} \cdot G \cdot \fder{\Sintd}{\dfield}
	-\frac{\eta}{2}
	\dfield \cdot \ep^{-1} \cdot \dfield
.
\label{eq:CanonicalERG}
\ee
An equation like~\eq{CanonicalERG} was first written down (without sources, but allowing for scale-dependence) in~\cite{Ball}. It can be thought of as a modification of Polchinski's equation in which the anomalous dimension of the fundamental field is explicitly taken into account; see~\cite{Fundamentals,HO-remarks} for detailed analyses.
A principle requirement for a valid ERG equation is that the kernels $G$ and $\ep^{-1}$---related via~\eq{Gb}---are quasi-local. Typically, $\ep$ is chosen according to~\eq{EffProp}, with the cutoff function conventionally normalized so that $\cutoff(0) = 1$ (further details can be found in~\cite{Fundamentals,HO-remarks}). Volume terms, discarded in this paper, are carefully treated in~\cite{HO-remarks}.

Deriving the analogous equation arising from special conformal transformations will be facilitated by the following. For some 
$V\bigl((x-y)^2\bigr)$, let us define
\be
	V_\mu(x,y) \equiv  (x+y)_\mu V\bigl((x-y)^2\bigr).
\label{eq:V_mu}
\ee

\begin{prop}
	Let $U(x,y) = U\bigl((x-y)^2\bigr)$ and suppose that, for some $V = V\bigl((x-y)^2\bigr)$,
	\[
		\dil{\Delta} U + U \dilL{\Delta} = V.
	\]
	Then it follows that
	\[
		\sct{\Delta}{\mu} U + U \sctL{\Delta}{\mu} = V_\mu.
	\]
	Proof: the result follows from the form of $\dil{\Delta}$ and $\sct{\Delta}{\mu}$ given in~\eqs{dil-qpo}{sct-qpo}.
\label{prop:dil-sct}
\end{prop}
Applying this result to~\eq{heq}, it is apparent that
\be
	\sct{\D-\delta+\eta}{\mu} h + h \sctL{\D-\delta+\eta}{\mu}
	=
	\eta \, {\ep^{-1}}_\mu,
\ee
where
\be
	{\ep^{-1}}_\mu(x,y) = (x+y)_\mu \ep^{-1}\bigl( (x-y)^2) \bigr)
.
\ee

Recalling~\eqss{paraK}{G_U}{auxK}, it is apparent that
\be
	\SCTR_{h\mu} = 
	\sct{\D-\delta}{\mu} \qsource \cdot \fder{}{\qsource}
	+
	e^{-\op} e^{-\hf \dfield \cdot h \cdot \dfield} 
	\biggl(
		\sct{\delta-\eta}{\mu} \dfield \cdot \fder{}{\dfield}
		-
		2\eta \, \partial_\mu \dfield \cdot \ep_0 \cdot \fder{}{\dfield}
	\biggr)
	e^{\hf \dfield \cdot h \cdot \dfield}  e^{\op}
.
\ee
Commuting $e^{\hf \dfield \cdot h \cdot \dfield}$ to the left, the generated of the form $\dfield \cdot \partial_\mu \ep_0 \cdot h \cdot \dfield $ vanishes due to the asymmetry of $\partial_\mu \ep_0 \cdot h$ under interchange of its arguments. In a little more detail we have, for some $H$:
\be
	\partial_\mu (\ep_0 \cdot h)
	\bigl(
		(x-y)^2
	\bigr)
	=
	(x-y)_\mu 
	H
	\bigl(
		(x-y)^2
	\bigr)
\ee
Now,
\be
	\Int{x}\volume{y}
	\dfield(x) (x-y)_\mu
	H\bigl(
		(x-y)^2
	\bigr)
	\dfield(y)
	=
	-
	\Int{x}\volume{y}
	\dfield(x) (x-y)_\mu
	H\bigl(
		(x-y)^2
	\bigr)
	\dfield(y)
	=0,
\ee
where in the second step we have swapped the dummy variables $x$ and $y$. Consequently,
we arrive at the following analogue of~\eq{Dil_h-prelim}:
\be
	\SCT_{h\mu}
	=
	\sct{\D-\delta}{\mu} \qsource \cdot \fder{}{\qsource}
	+
	e^{-\op}
	\biggl(
		\sct{\delta - \eta}{\mu} \dfield \cdot \fder{}{\dfield}
		-\frac{\eta}{2}
		\dfield
		\cdot
		{\ep^{-1}}_\mu
		\cdot
		\dfield
		-
		2 \eta \, \partial_\mu \dfield \cdot \ep_0 \cdot \fder{}{\dfield}
	\biggr)
	e^{\op}
.
\label{eq:SCT_h-prelim}
\ee
As before, the strategy is now to commute $e^{-\op}$ to the right.
To facilitate this, we note the following. First, recalling~\eq{Gb} and proposition~\ref{prop:dil-sct} it is apparent that
\be
	\biggl[
		e^{-\op}
		,
		\sct{\delta - \eta}{\mu} \dfield \cdot \fder{}{\dfield}
	\biggr] e^{\op}
	=
	\hf
	\fder{}{\dfield} \cdot G_\mu \cdot \fder{}{\dfield}
	+\frac{\eta}{2}
	 \fder{}{\dfield} \cdot \ep_\mu \cdot \fder{}{\dfield}	
.
\ee
Processing the next term in~\eq{SCT_h-prelim} gives:
\be
	-
	\biggl[
		e^{-\op}
		,
		\frac{\eta}{2}
		\dfield
		\cdot
		{\ep^{-1}}_\mu
		\cdot
		\dfield
	\biggr] e^{\op}
	=
	\eta
	\biggl(
	\dfield
	\cdot
	{\ep^{-1}}_\mu
	\cdot
	\ep
	\cdot
	\fder{}{\dfield}
	-
	\frac{1}{2}
	\fder{}{\dfield} \cdot \ep_\mu \cdot \fder{}{\dfield}
	\biggr)
,
\ee
where we have used the result
\be
	\ep \cdot {\ep^{-1}}_\mu \cdot \ep = \ep_\mu
.
\ee
This can be seen by reinstating arguments. Equivalently, note that the shorthand for~\eq{V_mu} is
$V_\mu = X_\mu V + V X_\mu$,  with $\bigl(X_\mu V\bigr)(x,y) = x_\mu V\bigl((x-y)^2\bigr)$ and $\bigl(VX_\mu\bigr)(x,y) = V\bigl((x-y)^2\bigr) y_\mu$, whereupon it follows that
\[
	\ep \cdot {\ep^{-1}}_\mu \cdot \ep
	=
	\ep \cdot X_\mu  \ep^{-1} \cdot \ep
	+
	\ep \cdot \ep^{-1} X_\mu \cdot \ep
	=
	\ep X_\mu + X_\mu \ep = \ep_\mu
.
\]

Finally, the last term in~\eq{SCT_h-prelim} gives, on account of translational invariance of $\ep_0$:
\be
	2 \eta
	\biggl[
		e^{-\op}
		,
		\partial_\mu \dfield \cdot \ep_0 \cdot \fder{}{\dfield}
	\biggr] e^{\op}
	=
	0
\ee
where, in accord with the discussion under~\eq{similarity}, equality strictly holds only up to a possible vacuum term.
We thus deduce that
\be
	\SCT_{h\mu}
	=
	\sct{\D-\delta}{\mu} \qsource \cdot \fder{}{\qsource}
	+
	\sct{\delta}{\mu} \dfield \cdot \fder{}{\dfield}
	+
	\hf
	\fder{}{\dfield} \cdot G_\mu \cdot \fder{}{\dfield} 	
	-\frac{\eta}{2}
	\dfield
	\cdot
	\bigl.\ep^{-1}\bigr._\mu
	\cdot
	\dfield
	+
	\eta \, \dfield \cdot f_\mu \cdot \fder{}{\dfield}
,
\label{eq:K_h}
\ee
where, noticing that the $\eta$ from the second term's $\delta - \eta$ has been pulled into the final term,
\be
	f_\mu = {\ep^{-1}}_\mu \cdot \ep + 2 \partial_\mu \ep_0 
	-  \identity X_\mu - X_\mu \identity
,
\ee
with $\identity(x,y) = \DD{\D}(x-y)$ implying that
\be
	 \bigl(
		\identity X_\mu + X_\mu \identity
	\bigr)
	(x,y)
	=
	(x+y)_\mu
	\DD{\D}(x-y)
.
\ee
We now recast $f_\mu$ in a simpler, manifestly quasi-local form. Recalling that $\ep = \ep_0 \cdot \cutoff$ and $\ep^{-1} = -\cutoff^{-1} \partial^2$ it follows that:
\begin{align*}
	{\ep^{-1}}_\mu \cdot \ep - X_\mu \identity 
	& =- \cutoff^{-1} \cdot \partial^2 X_\mu \, \ep_0 \cdot \cutoff
\\
	&= \cutoff^{-1} \cdot X_\mu \cutoff - 2 \cutoff^{-1} \cdot \partial_\mu \ep_0 \cdot \cutoff
\\
	&= \cutoff^{-1} \cdot X_\mu \cutoff - 2\partial_\mu \ep_0,
\end{align*}
where the last line follows from exploiting $\ep_0 \cdot \cutoff = \cutoff \cdot \ep_0$, together with
$\partial_\mu \cutoff \cdot \ep_0 = -\cutoff \cdot {\ep_0} \overleftarrow{\partial}_\mu$. Thus, we can simplify:
\be
	f_\mu = \cutoff^{-1} \cdot \bigl[ X_\mu, \cutoff\bigr]
.
\label{eq:f_mu}
\ee
However,
\begin{align}
	\pder{}{x_\alpha}
	\bigl(
		(x-y)_\alpha \cutoff
	\bigr)
	& =
	\bigl(\D + x\cdot \partial_x - y \cdot \partial_x\bigr) K
\nonumber
\\
	&=
	\bigl(\D + x\cdot \partial_x + y \cdot \partial_y\bigr) K
	= \partial^2 G
\end{align}
where, to go from the first line to the second, we have exploited translational invariance of $\cutoff$, with the final step following from~\eq{G}. From this, we deduce that
\be
	\bigl[
		X_\mu
	,
		\cutoff
	\bigr]
	=
	\partial_\mu G,
\ee
Therefore, the constraint on  the Wilsonian effective action implied by invariance under special conformal transformations, \eq{sctConstraint}, reads:
\begin{multline}
		\biggl(
			\sct{\D-\delta}{\mu}  \qsource
			\cdot \fder{}{\qsource} 
			+
			\sct{\delta}{\mu} \dfield \cdot \fder{}{\dfield} 
		\biggr)	
		\Sintd[\dfield,\qsource]
		=
\\
		\hf \fder{\Sintd}{\dfield} \cdot G_\mu \cdot  \fder{\Sintd}{\dfield} 
		- \hf \fder{}{\dfield}  \cdot G_\mu \cdot  \fder{\Sintd}{\dfield} 
		- \frac{\eta}{2} \dfield \cdot {\ep^{-1}}_\mu \cdot \dfield
		+\eta\, \partial_\mu \dfield \cdot \cutoff^{-1} \cdot G \cdot \fder{\Sintd}{\dfield}
.
\label{eq:CanonicalERG_mu}
\end{multline}
This equation is to the canonical ERG equation~\eq{CanonicalERG} what Sch{\"a}fer's equation~\cite{Schafer-Conformal} is to Wilson's ERG equation~\cite{Wilson}.

The generators in the ERG representation are constructed from~\eqs{D_h}{K_h} using the recipe in \eqs{actionD}{actionK}. The resulting expressions can be simplified by utilising the constraint equations~\eqs{CanonicalERG}{CanonicalERG_mu}.

\begin{representation} Canonical ERG representation
\begin{subequations}
\begin{align}
	\Trans_\mu & = \partial_\mu \qsource \cdot \fder{}{\qsource} + \partial_\mu \dfield \cdot \fder{}{\dfield},
\label{eq:Trans-Ball-rep}
\\
	\Rot_{\mu\nu} & =  L_{\mu\nu} \qsource \cdot \fder{}{\qsource} 
	+ L_{\mu\nu} \dfield \cdot \fder{}{\dfield}
\label{eq:Rot-Ball-rep}
\\
	\Dil & = \dil{\D-\delta} \qsource \cdot \fder{}{\qsource}
	+ D^{(\delta)} \dfield \cdot \fder{}{\dfield} 
	- \fder{\Sintd[\dfield,\qsource]}{\dfield}  \cdot G \cdot \fder{}{\dfield}
	+  \hf \fder{}{\dfield} \cdot G \cdot \fder{}{\dfield} 
,
\label{eq:Dil-Ball-rep}
\\
	\SCT_\mu & =  \sct{\D-\delta}{\mu} \qsource \cdot \fder{}{\qsource}
		+\sct{\delta}{\mu} \dfield \cdot \fder{}{\dfield} 
		-\fder{\Sintd[\dfield,\qsource] }{\dfield} \cdot G_\mu \cdot \fder{}{\dfield}
		+ \hf \fder{}{\dfield}  \cdot G_\mu \cdot \fder{}{\dfield}
,
\nonumber
\\
		&
		\qquad
		-\eta\,
		\partial_\mu \dfield \cdot \cutoff^{-1} \cdot G \cdot \fder{}{\dfield}
,
\label{eq:SCT-Ball-rep}
\end{align}
\end{subequations}
where $G$ and $G_\mu$ are defined in terms of $\ep$ via~\eqs{G}{V_mu}, the volume terms have been neglected and $\Sintd$ satisfies~\eqs{CanonicalERG}{CanonicalERG_mu}.
\label{rep:canonical}
\end{representation}

Though the analysis up to this point has been phrased in terms of conformal primary fields, we are at liberty to consider non-conformal theories: this can be done simply by taking the fields to which $\qsource_i$ couple as not being conformal primaries.

\section{The Energy-Momentum Tensor}
\label{sec:EM}

\subsection{Proposal}
\label{sec:proposal}

Given the scalar, conformal primary field, $\quasi^{(\delta)}$, we can furnish a representation of both this and a partner of scaling dimension $\D-\delta$ in terms of the appropriate sources:
\begin{subequations}
\begin{align}
	\quasi_{\qsource}^{(\delta)} & = \fder{\dual[\qsource]}{\qsource}
,
\label{eq:quasi^(Delta)}
\\
	\quasi_{\qsource}^{(\D-\delta)} & = \qsource
,
\label{eq:quasi^(D-Delta)}
\end{align}
\end{subequations}
where we recall that the subscript $\qsource$ denotes the Schwinger functional representation. It is immediately apparent that the pair of fields~\eqs{quasi^(Delta)}{quasi^(D-Delta)} satisfy \eq{qpo}. However, satisfaction of \eq{qpo} is a necessary but not sufficient condition for a field to belong to the spectrum of conformal primaries of a given theory. Indeed, we can construct any number of solutions to~\eq{qpo}, but only various combinations of solutions will correspond to the field contents of actual, realisable theories.

With this in mind, there are two assumptions at play in the statement that $\quasi^{(\delta)}$ and $\quasi^{(\D-\delta)}$ are conformal primaries. First, it is assumed that $\dual[\qsource]$ exists and is non-zero; we will encounter theories for which one or other of these conditions is violated in subsequent sections.
 More subtly, it is assumed that $\quasi^{(\D-\delta)}$ is amongst the spectrum of fields. As will be seen in \sect{qlrep}, if the ERG representation is quasi-local then $\quasi^{(\D-\delta)}$ is present in the spectrum as a redundant field.
 
Note that there are interesting theories for which the assumption that $\quasi^{(\D-\delta)}$ is in the spectrum of fields does not hold, in particular the mean field theories. This class of theories (recently featuring in e.g.~\cite{Rychkov-IsingI,Rychkov-IsingII,Heem+Pol-CFT,Fitz+Kaplan-Unitarity}) are such that the $n$-point functions are sums of products of two-point functions and cannot be represented in terms of a quasi-local action. The latter restriction amounts to defining mean field theories such as to exclude the Gaussian theory, plus its quasi-local but non-unitary cousins~\cite{Fundamentals} (see also \sect{non-unitary}); this is done for terminological convenience.
Thus, for mean field theories, \eq{quasi^(D-Delta)} amounts to minor notational abuse since, strictly, $\quasi$ should be reserved for conformal primaries. Accepting this, we henceforth interpret $\quasi_{\qsource}^{(\D-\delta)}$ as an object we are at liberty to construct, that in most---though not all---cases of interest is indeed a conformal primary. A surprising feature of mean field theories is that the energy-momentum tensor is not amongst the spectrum of conformal primary fields.%
\footnote{For a non-local two-point theory, $\eta/2$ is non-integer~\cite{Wegner-CS,Fundamentals}. With only two instances of the field and an even number of derivatives available, it is impossible to construct a local field of dimension, $\D$.
}

Sticking with the Schwinger functional representation, we construct a scalar field of scaling dimension $\D$:
\be
	\quasi_{\qsource}^{(\D)}
	= 
	-\delta \quasi_{\qsource}^{(\D-\delta)}\times \quasi_{\qsource}^{(\delta)}
	= 
	-\delta \qsource \times \fder{\dual[\qsource]}{\qsource}
,
\label{eq:mro-J}
\ee
where the factor of $-\delta$ is inserted so that, at least for theories satisfying the assumptions given above, we can identify $\quasi_{\qsource}^{(\D)}$ with the trace of the energy-momentum tensor.
The $\times$ symbol is just to emphasise that no integral is performed.
 For theories for which the Schwinger functional exists and is non-zero, but $\quasi^{(\D-\delta)}$ is not in the spectrum of the fields, we are again at liberty to construct $\quasi_{\qsource}^{(\D)}$, so long as we accept minor notional abuse and, more pertinently, that the energy-momentum tensor will not be amongst the spectrum of conformal primary fields. Recalling the discussion around~\eq{EMT-intro}, note that~\eq{mro-J}  is nothing but a statement of the Ward Identity corresponding to dilatation invariance of the Schwinger functional.

The above can be translated into a representation of our choice, though there is some subtlety in so doing. Leaving the choice of representation unspecified, let us tentatively write
\be
	\quasi^{(\D)} = -\delta \quasi^{(\D-\delta)} \times \quasi^{(\delta)}.
\label{eq:quasi^D-pre}
\ee
Restricting to the Schwinger functional representation clearly just recovers what we had before. However, there are representations---such as, crucially, the ERG representation---in which the dilatation and special conformal generators, \eqs{Dil-Ball-rep}{SCT-Ball-rep}, are second order in $\delta /\delta \dfield$. Acting with the dilatation generator, it is generally true that
\be
	\Dil \quasi^{(\D)} = 
	-\delta 
	\Bigl(
		\bigl[
			\Dil, \quasi^{(\D-\delta)}
		\bigr]	
		 \times \quasi^{(\delta)}
		 +
		 \quasi^{(\D-\delta)} \times \Dil \quasi^{(\delta)}
	 \Bigr)
.
\ee
For the Schwinger functional representation which is first order in functional derivatives, this reduces to
\be
	\Dil \quasi^{(\D)}_{\qsource} = 
	-\delta 
	\Bigl(
		\Dil \quasi^{(\D-\delta)}_{\qsource} 
		 \times \quasi^{(\delta)}_{\qsource} 
		 +
		 \quasi^{(\D-\delta)}_{\qsource}  \times \Dil \quasi^{(\delta)}_{\qsource} 
	 \Bigr)
	 =
	 \dil{\D} \quasi^{(\D)}_{\qsource} 
.
\ee
For the ERG representation, as will be seen explicitly in \sect{qlrep}, the solution is to extend $\quasi^{(\D-\delta)}_{\mathrm{loc}}$ to $\hat{\quasi}^{(\D-\delta)}_{\mathrm{loc}}$, with the latter such that
\be
	\bigl[
		\Dil, \hat{\quasi}^{(\D-\delta)}_{\mathrm{loc}}
	\bigr]	
	=
	\dil{\D-\delta} 
	\hat{\quasi}^{(\D-\delta)}_{\mathrm{loc}}
.
\label{eq:hatcommmutator}
\ee
With this in mind, we rewrite~\eq{quasi^D-pre} as
\be
	\quasi^{(\D)} = -\delta \hat{\quasi}^{(\D-\delta)} \times \quasi^{(\delta)},
\label{eq:quasi^D}
\ee
with the understanding that for `first order' representations, $\hat{\quasi}^{(\D-\delta)}$
just reduces to $\quasi^{(\D-\delta)}$.

Certain properties which are true of $\quasi^{(\D)}$ and its component fields are particularly transparent in the Schwinger functional representation. First of all, observe that, as discussed in the introduction, translation invariance implies:
\be
	\partial_\beta \quasi_{\qsource}^{(\D-\delta)} \cdot \quasi_{\qsource}^{(\delta)} = 0.
\label{eq:trans-inv}
\ee
Integrating by parts it follows that, for some $F_{\alpha\beta}$,
\be
	\quasi^{(\D-\delta)}_{\qsource} \times \partial_\beta \quasi^{(\delta)}_{\qsource}
	=
	-\partial_{\alpha} F_{\alpha\beta}.
\label{eq:F_ab}
\ee
Actually, in principle there could be an additional term which cannot be written as a total derivative but rather vanishes, when integrated, due to the integrand being odd. An example would be
\[
	\qsource(x) \Int{y} (x-y)_\beta \qsource(y).
\]
However, such terms are excluded if we insist that the theory in question possesses a quasi-local representation as we will do, henceforth. Recall that, in a quasi-local representation, all functions of the field have an expansion in positive powers of derivatives (it is blithely assumed that this expansion converges). For example, the derivative expansion of the action reads
\[
	\Sintd[\dfield] = 
	\Int{x} 
	\Bigl(
		V(\dfield) + Z(\dfield) (\partial_\mu \dfield)^2 + \cdots
	\Bigr)
\]
where $V(\dfield)$ is the local potential which, like $Z(\dfield)$, depends on $x$ only via the field (the ellipsis represent higher derivative terms).
With this in mind, let us consider~\eq{trans-inv} in a quasi-lcoal representation. Quasi-locality implies that any terms which vanish when integrated must take the form of total derivatives establishing that, for theories which permit a quasi-local formulation, \eq{F_ab} is correct as it stands.
Note that by focussing on theories supporting a quasi-local representation excludes mean field theories, in particular, from the remainder of the discussion.

We recognize that the form of~\eq{F_ab} is that of the Ward Identity associated with conservation of the energy-momentum tensor; inspired by this and~\eq{mro-J} we propose that for theories in which the energy-momentum tensor exists, \emph{the Ward Identities can be interpreted as defining a non-local representation of the energy momentum tensor}. Denoting the energy-momentum tensor in an arbitrary representation---which may or may not be quasi-local---by $\mathcal{T}_{\alpha \beta}$, we tentatively define this object via%
\footnote{Note that since the correlation functions involved in our proposed definition of the energy-momentum tensor involve only scalar fields, we expect symmetry under interchange of indices. 
}%
:
\begin{subequations}
\begin{align}
	\mathcal{T}_{\alpha \alpha}
	& = -\delta \hat{\quasi}^{(\D-\delta)} \times \quasi^{(\delta)}
,
\label{eq:traceT}
\\
	\partial_{\alpha} \mathcal{T}_{\alpha\beta}
	&=
	-\hat{\quasi}^{(\D-\delta)} \times \partial_\beta \quasi^{(\delta)}
,
\label{eq:divT}
\\
	\mathcal{T}_{\alpha\beta} 
	&= \mathcal{T}_{\beta\alpha}
\label{eq:symT}
.
\end{align}
\end{subequations}

\subsection{Justification}
\label{sec:Justify}

In this section, we justify, for $\D>1$, the proposal encapsulated in~\eqss{traceT}{divT}{symT}, which comprises three steps. First it is shown that an object which satisfies these equations is implied by translation, rotation and dilatation invariance so long as Polchinski's conditions~\cite{Pol-ScaleConformal} for the improvement of the energy-momentum tensor are satisfied. Secondly, it is shown how both the traceful and longitudinal components of $\mathcal{T}_{\alpha\beta}$ transform in a manner consistent with $\mathcal{T}_{\alpha\beta}$ being a conformal primary field of dimension $\D$. Finally,  the extent to which~\eqss{traceT}{divT}{symT} serve to uniquely define $\mathcal{T}_{\alpha\beta}$ is discussed.

\subsubsection{Existence}
\label{sec:Existence}

The most basic requirement for the existence of a non-null $\mathcal{T}_{\alpha\beta}$, as defined via~\eqss{traceT}{divT}{symT}, is that the fields $\hat{\quasi}^{(\D-\delta)}$ and $\quasi^{(\delta)}$ exist and are non-zero. There is some degree of subtelty here since it is conceivable that 
 $\quasi^{(\delta)}$ does not exist in the Schwinger functional representation but does exist in a quasi-local representation. An example would be the Gaussian theory in $\D=2$. The IR behaviour of this theory is sufficiently bad that the lowest dimension conformal primary is not $\dfield$ but rather $\partial_\mu \dfield$. One method for dealing with this theory would be to perform the analysis of this section in terms of vector fields. An alternative, however, is to implicitly work within a quasi-local representation; note that though $\quasi^{(\delta)}$ exists, we are accepting a degree of notational abuse since it is not a conformal primary in the standard sense. (Later, where more care must be taken, the symbol $\phi$ used, instead).
With this in mind, for the duration of this section we assume that at least one representation of $\hat{\quasi}^{(\D-\delta)}$ and $\quasi^{(\delta)}$ exists, and at least one of these representations is quasi-local.

Given this, we now move on to determining the conditions under which~\eqss{traceT}{divT}{symT} are implied by a combination of translation, rotation and dilatation invariance. 
In \sect{proposal}, we have already seen that, for some $F_{\alpha \beta}$, translation invariance plus the existence of a quasi-local representation implies~\eq{F_ab}. Similarly, from rotational invariance it follows that
\be
	\Int{x} \hat{\quasi}^{(\D-\delta)}(x) \bigl(x_\alpha \partial_\beta - x_\beta \partial_\alpha \bigr) \quasi^{(\delta)}(x) = 0.
\ee
Substituting in~\eq{F_ab} we have:
\be
	\Int{x}
	\bigl(
		F_{\alpha \beta}(x) - F_{\beta \alpha}(x)
	\bigr)
	=
	0,
\ee
implying that, for some $f_{\lambda \alpha\beta}$, antisymmetric in its last two indices,
\be
	F_{\alpha \beta} - F_{\beta \alpha} = \partial_\lambda f_{\lambda \alpha\beta}
.
\label{eq:F_ab-F_ba}
\ee
We might wonder whether, as in the case of translation invariance, an additional term can appear on the \rhs\ that cannot be expressed as a total derivative. This would be of the form $Y_{\alpha \beta} - Y_{\beta \alpha}$, where $\Int{x} Y_{\alpha\beta}$ is symmetric. Again, quasi-locality guarantees that such contributions can in fact be absorbed into the total derivative term, $\partial_\lambda f_{\lambda \alpha\beta}$.

Inspired by the standard derivation of the Belinfante tensor we observe that, for some $F_{\lambda\alpha\beta}$ antisymmetric in its \emph{first} two indices, \eq{F_ab} is left invariant by the shift
\be
	F_{\alpha \beta} \rightarrow F_{\alpha \beta} - \partial_{\lambda} F_{\lambda\alpha\beta}
.
\label{eq:shift}
\ee
Under this transformation, \eq{F_ab-F_ba} becomes:
\be
	F_{\alpha \beta} - F_{\beta \alpha} 
	= 
	\partial_\lambda 
	\bigl(
		f_{\lambda \alpha\beta}
		+ F_{\lambda \alpha \beta}  
		- F_{\lambda \beta \alpha}
	\bigr)
,
\label{eq:F_ab-Prelim}
\ee
and so we choose $F_{\lambda \alpha \beta}$ such that
\be
	F_{\alpha \beta} - F_{\beta \alpha} 
	= 0.
\label{eq:F_ab-Sym}
\ee
Note that the choice of $F_{\lambda \alpha \beta}$ is not unique: we can add to it further terms, antisymmetric in the first pair of indices and symmetric in the second, which leave both~\eqs{F_ab}{F_ab-Sym} invariant. 

Let us summarize what has been accomplished so far. On the one hand we have demonstrated the existence of a tensor field, $F_{\alpha \beta}$, which satisfies:
\begin{subequations}
\begin{align}
	\partial_{\alpha} F_{\alpha\beta}
	&=
	-\hat{\quasi}^{(\D-\delta)} \times \partial_\beta \quasi^{(\delta)}
,
\label{eq:divF}
\\
	F_{\alpha\beta} 
	&= F_{\beta\alpha}
.
\label{eq:symF}
\end{align}
\end{subequations}
On the other hand, we have shown the existence of a scalar field, $\mathcal{T}_{\alpha\alpha}$, satisfying~\eq{traceT}. Next we must show that these two objects are, in fact, related. To do this, we start by multiplying~\eq{divF} by $x_\beta$ and integrating over all space, giving
\be
	\int F_{\alpha \alpha} = 
	- \dil{\D-\delta} \hat{\quasi}^{(\D-\delta)} \cdot \quasi^{(\delta)}
	- \delta \hat{\quasi}^{(\D-\delta)} \cdot \quasi^{(\delta)}
.
\ee
The first term on the \rhs\ vanishes as a consequence of dilatation invariance and so we conclude that, for some $f_\lambda$,
\be
	\mathcal{T}_{\alpha \alpha}  = F_{\alpha \alpha} + \partial_\lambda f_\lambda
.
\label{eq:F-T}
\ee
But, as noted above, \eqs{divF}{symF} are left invariant by the shift
\be
	F_{\alpha \beta} \rightarrow F_{\alpha\beta} - \partial_\lambda Z_{\lambda \alpha \beta},
\label{eq:shiftTrace}
\ee
where
\be
	\partial_\lambda Z_{\lambda \alpha \alpha} = \partial_\lambda f_\lambda,
	\qquad
	Z_{\lambda \alpha \beta} = - Z_{\alpha \lambda \beta},
	\qquad
	\partial_\lambda Z_{\lambda \alpha \beta} = \partial_\lambda Z_{\lambda \beta \alpha}
.
\label{eq:Z_lab}
\ee
In $\D=1$ it is impossible to construct a non-zero $Z_{\lambda \beta \alpha}$. For $\D>1$, it would appear that we can eliminate the unwanted term via the shift~\eq{shiftTrace}, while leaving~\eqs{divF}{symF} invariant. However, there is a subtlety.

Let us observe that, apparently, for $\D > 1$ a solution to the constraints on $Z_{\lambda \alpha \beta}$ always exists:
\be
	\partial_\lambda Z_{\lambda \alpha \beta}
	=
	\frac{1}{1-\D}
	\bigl(
		\delta_{\alpha \beta} \partial^2 - \partial_\alpha \partial_\beta
	\bigr)
	\ep_0 \cdot \partial_\lambda f_\lambda
.
\label{eq:non-local}
\ee
The transverse structure guarantees that the result of contracting with either $\partial_\alpha$ or $\partial_\beta$ vanishes, as required.
However, the theories under examination are required to support a quasi-local representation. Since $\partial_\alpha \partial_\beta \ep_0$ is non-local, the above solution is not always valid. If $f_\lambda$ is a descendant field of the form $\partial_\lambda \quasi$ or $\partial^2 \mathrm{\quasi_\lambda}$ then there is no problem since the non-locality is ameliorated. Interestingly, there is an additional possibility: it could be that $f_\lambda$ can be written as $\partial_\lambda \phi$, where here and below we understand $\phi$ as being neither a primary nor descendant field. This may seem exotic but recall that, for the Gaussian fixed-point in $\D=2$, the fundamental field is just such a scalar; indeed, in this case the lowest dimension primary field is $\partial_\lambda \phi$.

Excluding this exotic case, it is clear that the solution~\eq{non-local} is no good if $f_\lambda$ is a primary field. Moreover, it fails for the case that $f_\lambda = \partial_\sigma f_{\sigma \lambda}$ for some $f_{\sigma \lambda}$ that does not reduce to $\delta_{\sigma\lambda}$. In $\D=2$, this is the end of the story, but this is not so for higher dimensionality. Let us suppose that, for some $f_{\sigma \lambda}$
\be
	f_\lambda = \partial_\sigma f_{\sigma \lambda}.
\ee
If $ f_{\sigma \lambda}$ contains the Kronecker-$\delta$ then we write $f_{\sigma\lambda} = \delta_{\sigma\lambda} f$.
The quasi-local solutions are:
\begin{subequations}
\begin{align}
\nonumber
	\partial_\lambda Z_{\lambda \alpha \beta}
	& =
	\frac{1}{2-\D}
	\bigl(
		\partial_\alpha \partial_\sigma f_{\sigma \beta} +
		\partial_\beta \partial_\sigma f_{\sigma \alpha}
		-\partial^2 f_{\alpha \beta}
		-\delta_{\alpha \beta} \partial_\sigma \partial_\lambda f_{\sigma \lambda} 
	\bigr)
\\
	&
	\qquad+
	\frac{1}{(2-\D)(\D-1)}
	\bigl(
		\delta_{\alpha \beta} \partial^2
		-\partial_{\alpha} \partial_\beta
	\bigr) f_{\sigma \sigma}
	&
	\mathrm{for}\ \D > 2
,
\label{eq:Pol-condition-d>2}
\\
	\partial_\lambda Z_{\lambda \alpha \beta}
	& =
	\frac{1}{1-\D}
	\bigl(
		\partial_{\alpha} \partial_\beta
		-\delta_{\alpha \beta} \partial^2
	\bigr) f
	&
	\mathrm{for}\ \D = 2	
.
\label{eq:Pol-condition-d=2}
\end{align}
\end{subequations}
Observe that these conditions correspond to the recipe found by Polchinski~\cite{Pol-ScaleConformal} for improving the energy-momentum tensor of a conformal field theory such that it is traceless.

With these points in mind, consider the sufficient conditions for translation, rotation and dilatation invariance to imply~\eqss{traceT}{divT}{symT}. In $\D>2$, absence of a primary vector field of scaling dimension $\D-1$ is sufficient. In $\D=2$ this condition must be supplemented by at the absence of vector fields that can be written as $\partial_\sigma f_{\sigma \lambda}$, where $f_{\sigma \lambda}$ does not reduce to the Kronecker-$\delta$. In some sense, this is academic since it cannot be realised for unitary theories. Moreover,  as we shall see in \sect{Conditions}, this additional condition is relevant only for theories with sufficiently bad IR behaviour.

To conclude this section, let us mention an alternative way to recover Polchinksi's conclusion as to the conditions under which improvement of the energy-momentum tensor is possible. A version of the analysis below forms a key part of~\cite{Delamotte-Conformal}, in which an argument is given as to why scale invariance is automatically enhanced to conformal invariance for the Ising model in three dimensions.

Using the `auxiliary functional representation' of the conformal algebra: \eq{auxP}--\eq{auxK}, consider~\eqs{dilConstraint}{sctConstraint},  supposing that conformal invariance is yet be established but scale invariance holds:
\begin{subequations}
\begin{align}
	\SCTR_{\mathcal{U} \mu} e^{-\Sintd[\dfield, \qsource]}
	&=  E_{\Sintd \mu }[\dfield, \qsource] e^{-\Sintd[\dfield, \qsource]},
\label{eq:E_mu}
\\
	\DilR_{\mathcal{U}} e^{-\Sintd[\dfield, \qsource]} & = 0
.
\end{align}
\end{subequations}
Acting on the first equation with $\DilR_{\mathcal{U}} $ and exploiting the second equation, together with $[\DilR_{\mathcal{U}}, \SCTR_{\mathcal{U} \mu}] = -\SCTR_{\mathcal{U} \mu}$ yields:
\be
	\bigl[
		\DilR_{\mathcal{U}}, E_{\Sintd \mu }
	\bigr]
	= - E_{\Sintd \mu }
.
\label{eq:Dil:E_mu}
\ee
Invariance of $E_{\Sintd \mu }$ under translations and rotation follows by acting on~\eq{E_mu} with $\Trans_\mu$ and $\Rot_{\mu\nu}$ and exploiting invariance of $\Sintd$ under translations, rotations and dilatations. Thus, as claimed in~\cite{Delamotte-Conformal}, \eq{Dil:E_mu} implies that $E_{\Sintd \mu }[\dfield, \qsource]$ must be expressible as a combination of integrated fields of dimension $\D-1$. Therefore, scale invariance implies conformal invariance if either there are no primary vector fields of scaling dimension $\D-1$ or any such fields can be expressed as $\partial_\mu \phi$.

\subsubsection{Conformal Covariance}
\label{sec:ConformalCovariance}

In this section we will show that~\eqss{traceT}{divT}{symT} are consistent with $\mathcal{T}_{\alpha\beta}$ being a candidate for a conformal primary field of dimension $\D$. By `candidate field' we mean an object that has the desired properties under conformal transformations but may or may not turn out to be amongst the spectrum of conformal primaries for a given theory. Let us start by noting that, by construction, $\mathcal{T}_{\alpha \alpha}$ is a candidate for a conformal primary field (\cf~\eq{mro-J}). The rest of this section will be devoted to showing that the longitudinal parts of $\mathcal{T}_{\alpha \beta}$ also transform correctly. 

As a warm up, first let us confirm translational covariance of $\mathcal{T}_{\alpha\alpha}$.
Operating on~\eq{traceT} with the generator of translations, $\Trans_{\mu}$:
\be
	\Trans_{\mu} \mathcal{T}_{\alpha\alpha} 
	= -\delta 
	\Bigl(
		\partial_\mu \hat{\quasi}^{(\D-\delta)} \times  \quasi^{(\delta)}
		+
		\hat{\quasi}^{(\D-\delta)} \times  \partial_\mu \quasi^{(\delta)}
	\Bigr)
	=
	\partial_\mu \mathcal{T}_{\alpha\alpha} 
,
\ee
where we have exploited~\eq{qpo}. Therefore, as expected, $\mathcal{T}_{\alpha \alpha}$ is translationally invariant. We can now play a similar game with~\eq{divT} to show that $\partial_\alpha \mathcal{T}_{\alpha\beta}$ is translationally invariant.
In the same vein, it is straightforward, by considering the action of the dilatation generator, to show that both $\mathcal{T}_{\alpha\alpha}$ and the longitudinal components of $ \mathcal{T}_{\alpha \beta}$ have scaling dimension, $\D$.

Covariance of $\mathcal{T}_{\alpha \alpha}$ under rotations and special conformal transformations follows exactly the same pattern. Dealing with~\eq{divT} is only ever so slightly more involved. To start with, let us consider rotations:
\begin{align}
	\Rot_{\mu\nu} \partial_\alpha \mathcal{T}_{\alpha \beta}
	& 
	=
	-\rot{\mu\nu} \hat{\quasi}^{(\D-\delta)} \partial_\beta \quasi^{(\D)} 
	- \hat{\quasi}^{(\D-\delta)}\partial_\beta \rot{\mu\nu}  \quasi^{(\D)}
\nonumber
\\
	&
	= \rot{\mu\nu} \partial_\alpha \mathcal{T}_{\alpha \beta}
	+\delta_{\mu\beta} \partial_\alpha \mathcal{T}_{\alpha \nu}
	-\delta_{\nu\beta} \partial_\alpha \mathcal{T}_{\alpha \mu}
.
\label{eq:M-divT}
\end{align}
Now, given a symmetric conformal primary tensor field, $\quasi^{(\D)}_{\alpha\beta}$, let us compare~\eq{M-divT} with
$\partial_\alpha \quasi^{(\D)}_{\alpha\beta}$, the result of which we can compute using~\eq{rot-qpo-tensor}:
\begin{align}
	\Rot_{\mu\nu} \partial_\alpha \quasi^{(\D)}_{\alpha\beta}
&	=
	\partial_\alpha 
	\Bigl(
		\rot{\mu\nu} \quasi^{(\D)}_{\alpha\beta}
		+
		\bigl(
			\delta_{\mu\alpha} \delta_{\lambda \nu} - \delta_{\nu\alpha} \delta_{\lambda\mu}
		\bigr)		
		\quasi^{(\D)}_{\lambda \beta}
		+
		\bigl(
			\delta_{\mu\beta} \delta_{\lambda \nu} - \delta_{\nu\beta} \delta_{\lambda\mu}
		\bigr)		
		\quasi^{(\D)}_{\alpha \lambda}
	\Bigr)
\nonumber
\\
&	=
	\rot{\mu\nu} \, \partial_\alpha \quasi^{(\D)}_{\alpha\beta}
	+\delta_{\mu\beta} \partial_\alpha \quasi^{(\D)}_{\alpha \nu}
	-\delta_{\nu\beta} \partial_\alpha \quasi^{(\D)}_{\alpha \mu}
.
\label{eq:M-divO}
\end{align}
Comparing~\eqs{M-divT}{M-divO}, we conclude that the longitudinal pieces of $\mathcal{T}_{\alpha\beta}$ transform under rotations like a conformal primary tensor field.

Finally, we deal with special conformal transformations:
\begin{align}
	\SCT_{\mu} \partial_\alpha \mathcal{T}_{\alpha \beta}
	& 
	=
	-\sct{\D-\delta}{\mu} \hat{\quasi}^{(\D-\delta)} \partial_\beta \quasi^{(\delta)} 
	- \hat{\quasi}^{(\D-\delta)} \partial_\beta \sct{\delta}{\mu}  \quasi^{(\delta)}
\nonumber
\\
	&
	= \sct{\D}{\mu} \partial_\alpha \mathcal{T}_{\alpha\beta}
	+ 2\delta_{\mu\beta}
	\bigl( 
		x_\lambda \partial_\alpha \mathcal{T}_{\alpha\lambda}
		+ \mathcal{T}_{\alpha\alpha}
	\bigr)
	+2 x_\mu \partial_\alpha \mathcal{T}_{\alpha \beta}
	- 2 x_\beta \partial_\alpha \mathcal{T}_{\alpha \mu}
.
\label{eq:K-divT}
\end{align}
Now, given a symmetric conformal primary tensor field, $\quasi^{(\D)}_{\alpha\beta}$, let us compare~\eq{K-divT} with
$\SCT_{\mu} \partial_\alpha \quasi^{(\D)}_{\alpha\beta}$, the result of which we can compute using~\eq{sct-qpo-tensor}. Exploiting symmetry under $\alpha \leftrightarrow \beta$ it is straightforward to show that
\be
	\SCT_{\mu} \partial_\alpha \quasi^{(\D)}_{\alpha \beta}
	=
	\sct{\D}{\mu} \partial_\alpha \quasi^{(\D)}_{\alpha\beta}
	+ 2\delta_{\mu\beta}
	\bigl( 
		x_\lambda \partial_\alpha \quasi^{(\D)}_{\alpha \lambda}
		+ \quasi^{(\D)}_{\alpha\alpha}
	\bigr)
	+2 x_\mu \partial_\alpha \quasi^{(\D)}_{\alpha \beta}
	- 2 x_\beta \partial_\alpha \quasi^{(\D)}_{\alpha \mu}
.
\label{eq:K-divO}
\ee
Comparing~\eqs{K-divT}{K-divO} it is apparent that the longitudinal pieces of $\mathcal{T}_{\alpha\beta}$ transforms under conformal transformations like a conformal primary tensor field of dimension $\D$. 

\subsubsection{Uniqueness}

A two-index tensor has $\D^2$ \emph{a priori} independent components. The condition of symmetry under interchange of indices imposes $\D(\D-1)/2$ constraints; conservation imposes a further $\D$, whereas~\eq{traceT} yields one additional constraint. This reduces the number of independent components to
\[
	\frac{(\D-2)(\D+1)}{2}
.
\]
Immediately it is apparent that~\eqss{traceT}{divT}{symT} uniquely define the energy-momentum tensor in $\D=2$. 

For $\D>2$, we must accept that, in general, $\mathcal{T}_{\alpha \beta}$ is not uniquely defined.
Notice that the equations~\eqss{traceT}{divT}{symT} are invariant under
\be
	\mathcal{T}_{\alpha\beta} \rightarrow \mathcal{T}_{\alpha\beta} + Z_{\alpha\beta},
\ee
where
\be
	Z_{\alpha \alpha} = 0,
	\qquad
	\partial_{\alpha} Z_{\alpha \beta} = 0,
	\qquad
	Z_{\alpha\beta} = Z_{\beta\alpha}
.
\label{eq:unique}
\ee

The results of the previous sub-section show that the traceful and longitudinal components of $\mathcal{T}_{\alpha \beta}$ transform as expected for a conformal primary of dimension $\D$.
Let us now focus on conformal field theories for which the energy-momentum tensor exists.  We assume that $Z_{\alpha \beta}$ is chosen such that any remaining components of $\mathcal{T}_{\alpha \beta}$ also transform homogeneously. However, this still leaves a residual freedom to add to  $Z_{\alpha \beta}$ a contribution, $\tilde{Z}_{\alpha \beta}$, which also transforms like a conformal primary of dimension $\D$. This requirement, together with~\eq{unique}, implies that~\cite{HO-EMT}:
\be
	\tilde{Z}_{\alpha \beta} = \partial_\rho \partial_\sigma C_{\alpha \rho \beta \sigma},
\ee
where $C_{\alpha \rho \sigma \beta}$ is a conformal primary field of dimension $\D-2$ with the same symmetries as the Weyl tensor: \be
\begin{split}
	 C_{\alpha \rho \beta \sigma } =  -C_{\rho\alpha \beta \sigma} = - C_{\alpha \rho \sigma \beta} ,
\\
	  C_{\alpha \rho \beta \sigma } + C_{\alpha \beta \sigma \rho} + C_{\alpha \sigma \rho \beta } = 0,
\\
	  C_{\alpha \rho \alpha \sigma} = 0.
\end{split}
\ee
For $\D=3$, these constraints do not have a  non-trivial solution and so extend the uniqueness of the energy-momentum, for a conformal field theory, to this dimensionality. Beyond this, uniqueness or otherwise depends on whether or not the theory in question supports $C_{\alpha \rho \beta \sigma }$ as a conformal primary field of dimensions $\D-2$~\cite{HO-EMT}.

Though we will not rely on the following restriction in this paper, it is expected that the energy-momentum tensor is unique for unitary theories.%
\footnote{I would like to thank H.~Osborn for informing me of this and for providing the argument as to why.}
In $\D=4$, $C_{\alpha \rho \beta \sigma}$ transforms under the $(2,0) \oplus (0,2)$ representation. However, Mack rigorously established that for a representation of type $(j,0)$, unitarity demands that the scaling dimension $\Delta > 1 + j$~\cite{Mack-Unitary}. This implies that the scaling dimension of $\partial_\rho \partial_\sigma C_{\alpha \rho \beta \sigma}$ is greater than five and so this field cannot contribute to the energy-momentum tensor which, in the considered dimensionality, is of scaling dimension four. A similar result is expected to hold in higher dimensions.

\subsection{Conformal Invariance}
\label{sec:Conditions}

We have previously established the conditions under which~\eqss{traceT}{divT}{symT} hold. Given these equations, it is a simple matter to demonstrate conformal invariance (indeed, we have essentially shown that the energy-momentum tensor can be improved to a traceless, symmetric form). Recall that the condition for conformal invariance reads:
\be
	\sct{\D-\delta}{\mu} \qsource
	\cdot
	\fder{\dual[\qsource]}{\qsource}
	= 0
\ee
which, in an arbitrary representation, becomes:
\be
	\sct{\D-\delta}{\mu} \hat{\quasi}^{(\D-\delta)}
	\cdot 
	\quasi^{(\delta)} = 0,
	\qquad
	\Rightarrow
	\qquad
	\hat{\quasi}^{(\D-\delta)}
	\cdot 
	\sct{\delta}{\mu}
	\quasi^{(\delta)} 
	= 0.
\label{eq:SC-arbitrary}
\ee
With this in mind, consider a theory for which full conformal invariance is yet to be established. Utilizing~\eqss{traceT}{divT}{symT}, we see that
\begin{align}
	\hat{\quasi}^{(\D-\delta)}
	\cdot 
	\sct{\delta}{\mu} \quasi^{(\delta)} 
&
=
	\hat{\quasi}^{(\D-\delta)}
	\cdot 
	\bigl(
		2x_\mu (x\cdot \partial + \delta) -x^2 \partial_\mu
	\bigr)
	\quasi^{(\delta)}
\nonumber
\\
&
=
	\Int{x} 
	\bigl(
		x^2 \partial_\alpha \mathcal{T}_{\alpha\mu}
		-2x_\mu x_\beta \partial_\alpha \mathcal{T}_{\alpha \beta}	
		-2x_\mu \mathcal{T}_{\alpha\alpha}
	\bigr)
\nonumber
\\
&= 	2 \Int{x} x_\alpha
	\bigl(
		\mathcal{T}_{\mu \alpha} - \mathcal{T}_{\alpha \mu} 
	\bigr)
\nonumber
\\
&=
0
\label{eq:ImpliedConformal}
\end{align}
where, to go from the second to third lines we have integrated by parts, and to go to the last line we have exploited symmetry of $\mathcal{T}_{\mu \alpha}$ under interchange of indices.  Therefore, conformal invariance has been demonstrated.

It is interesting to consider theories for which the energy-momentum tensor cannot be improved to be traceless. According to~\eq{F-T}, there is a residual term of the form $\partial_\lambda f_\lambda$ which spoils conformal invariance. Repeating the analysis of~\eq{ImpliedConformal}, it is apparent that
\be
	\hat{\quasi}^{(\D-\delta)}
	\cdot 
	\sct{\delta}{\mu} \quasi^{(\delta)} 
	\propto
	\Int{x}
	f_\mu
.
\label{eq:NoConformal}
\ee
The intriguing thing about this condition is that, in $\D=2$, the sufficient conditions for the improvement of the energy-momentum tensor include the absence of a vector field which can be written as $\partial_\sigma f_{\sigma \lambda}$, where $f_{\sigma \lambda}$ does not reduce to the Kronecker-$\delta$. However, suppose that $f_\mu$ can be written in this fashion; according to~\eq{NoConformal}, conformal invariance is present since the \rhs\ vanishes! This apparent paradox is resolved by noting that, in the quasi-local representation appropriate to the discussion of the improvement 
energy-momentum tensor, 
$
	\hat{\quasi}^{(\D-\delta)}
	\cdot 
	\sct{\delta}{\mu} \quasi^{(\delta)} 
	= 0
$
does not quite imply conformal invariance of the correlation functions: this only holds if $\dual[\qsource]$ exists, which it will not if the IR behaviour is sufficiently bad. This will be exemplified in \sect{non-unitary} by a non-unitary theory.

\subsection{Quasi-Local Representation}
\label{sec:qlrep}

The defining equations for the energy-momentum tensors, \eqss{traceT}{divT}{symT} are independent of any particular representation. As already apparent, a prominent role is played by the Schwinger functional representation; the other representation of particular interest is furnished by the ERG, which provides a quasi-local framework. For the sake of definiteness, in this section we will explore the energy-momentum tensor using the canonical ERG equation, discussed in \sect{Pol}. 
Recall that $\delta$ can be written in terms of the anomalous dimension of the fundamental via~\eq{delta_star}.

In this context, we have:
\begin{subequations}
\begin{align}
	\quasi^{(\delta)}_{\mathrm{loc}}
	&
		=  \cutoff^{-1} \cdot \bigl(1+\varrho\bigr) \cdot \dfield 
		+ \ep_0 \cdot \varrho \cdot \fder{\Sintd}{\dfield}
		,
\label{eq:quasi_delta-local}
\\
	\quasi^{(\D-\delta)}_{\mathrm{loc}} & 
	=  \dfield \cdot  \ep^{-1}_0 +  \fder{\Sintd}{\dfield} \cdot \cutoff
	,
\end{align}
\end{subequations}
where, in momentum space, for $\eta <2$
\be
	\varrho(p^2) = -p^{2(\eta/2)} \cutoff(p^2)
	\int_0^{p^2} d q^2
	\biggl[
		\frac{1}{\cutoff(q^2)}
	\biggr]'
	q^{-2(\eta/2)}
,
\label{eq:rho}
\ee
with $\cutoff$ the momentum-space cutoff introduced in~\eq{EffProp}. The expressions for $\varrho$ and $\quasi^{(\delta)} $ first appeared in~\cite{HO+JOD} and played a prominent role in much of the analysis of~\cite{Fundamentals}. The pair of fields~\eqs{quasi_delta-local}{quasi_delta-local}
appears in both~\cite{HO-remarks,Fundamentals}. Note that the $h$ appearing in~\eq{W_h} is (given appropriate boundary conditions) related to $\varrho$ according to~\cite{Fundamentals}:
\be
	h = \ep^{-1} \cdot \bigl( \one + \varrho \bigr) + h_0,
	\qquad
	h_0 = 
	\biggl\{
	\begin{array}{cl}
		1, & \eta = 0,
	\\
		0, & \eta < 2, \neq 0.
	\end{array}
\ee

For our purposes, though, we seek a slight modification to $\quasi^{(\D-\delta)}_{\mathrm{loc}}$,
as discussed around \eq{quasi^D-pre}.
To motivate this, consider~\eq{quasi^D}. The novelty of the current representation is that the dilatation generator~\eq{Dil-Ball-rep} has a term containing not one but two functional derivatives. Consequently,
\[
	\DilR_{\Sintd}[\dfield] \operator \neq 
	\bigl[
		\DilR_{\Sintd}[\dfield], \operator
	\bigr]
.
\]
Therefore, we seek an extension of $\quasi^{(\D-\delta)}_{\mathrm{loc}}$, to be denoted $\hat{\quasi}^{(\D-\delta)}_{\mathrm{loc}}$, such that
\be
	\bigl[
		\DilR_{\Sintd}[\dfield], 
		\hat{\quasi}^{(\D-\delta)}_{\mathrm{loc}}
	\bigr]
	=
	\dil{\D-\delta}
	\hat{\quasi}^{(\D-\delta)}_{\mathrm{loc}}
.
\label{eq:extension}
\ee
In this way we can construct
\be
	\quasi^{(\D)}_{\mathrm{loc}}
	= 
	-\delta \hat{\quasi}^{(\D-\delta)}_{\mathrm{loc}} \times
	\quasi^{(\delta)}_{\mathrm{loc}}
,
\ee
which by construction satisfies
\be
	\DilR_{\Sintd}[\dfield]
	\quasi^{(\D)}_{\mathrm{loc}}
	=
	\dil{\D}
	\quasi^{(\D)}_{\mathrm{loc}}
.
\ee
The solution is to take
\be
	\hat{\quasi}^{(\D-\delta)}_{\mathrm{loc}}
	=  \dfield \cdot  \ep^{-1}_0 +  \fder{\Sintd}{\dfield} \cdot \cutoff
	-  \fder{}{\dfield} \cdot \cutoff
.
\label{eq:quasi_d-delta-local}
\ee

There a various ways to obtain this equation. On the one hand, a brute force calculation can be performed, along the lines of appendix C of~\cite{Fundamentals}. However, there is a more elegant approach. Notice that we may write
\be
	\hat{\quasi}^{(\D-\delta)}_{\mathrm{loc}}
	=
	e^{\Sintd} e^{-\op} \, \dfield \cdot  \ep^{-1}_0 \, e^{\op} e^{-\Sintd}
.
\ee
Recalling~\eqs{gen_S}{Dil_h-prelim}, it is apparent that
\be
	\Bigl[
		\DilR_{\Sintd}[\dfield],
		\hat{\quasi}^{(\D-\delta)}_{\mathrm{loc}}
	\Bigr]
	=
	e^{\Sintd} e^{-\op} \, \dil{\delta-\eta} \dfield \cdot  \ep^{-1}_0 \, e^{\op} e^{-\Sintd}
.
\ee
Splitting $\delta - \eta = \delta_0 - \eta/2$ and noting that
\be
	\dil{\D-\delta_0} \ep^{-1}_0 +  \ep^{-1}_0 \dilL{\D-\delta_0} = 0,
\ee
it follows that
\begin{align}
	\dil{\delta-\eta} \dfield \cdot  \ep^{-1}_0
	& =
	-\frac{\eta}{2} \dfield \cdot  \ep^{-1}_0
	-\dfield \cdot \dil{\D-\delta_0} \ep^{-1}_0
\nonumber
\\	
	& =
	-\frac{\eta}{2} \dfield \cdot  \ep^{-1}_0 + \dfield \cdot \ep^{-1}_0 \dilL{\D - \delta_0}	
\nonumber
\\
	&=
	\dfield \cdot \ep^{-1}_0 \dilL{\D - \delta}
,
\end{align}
confirming~\eq{extension}.

As we know from~\eq{traceT}, the fields discussed above can be combined to form the trace of the energy-momentum tensor. Integrating over space, we recognize the resulting object as nothing other than the exactly marginal, redundant field which exists at every critical fixed-point. Recall that the definition of a redundant field is that it can be cast as a quasi-local field redefinition. This is essentially manifest in the case of the trace of the energy-momentum tensor---which in the ERG representation we denote by $T_{\alpha\alpha}$---since
\begin{align}
	\int T_{\alpha\alpha}
	&=
	-\delta \hat\quasi^{(\D-\delta)}_{\mathrm{loc}} \cdot \quasi^{(\delta)}_{\mathrm{loc}}
\nonumber
\\
	&=
	-\delta e^{\Sintd[\dfield]}
	\biggl(
		\fder{}{\dfield} - \dfield \cdot \ep^{-1}
	\biggr)
	\cdot
	\biggl(
		\ep \cdot \varrho \cdot \fder{}{\dfield}
		- \bigl(1+\varrho\bigr) \cdot \dfield 
	\biggr)
	e^{-\Sintd[\dfield]}
.
\end{align}
Recalling~\eq{TotalAction} we have, for infinitesimal $\epsilon$:
\be
	e^{-\Stotd[\dfield] + \epsilon \int T_{\alpha\alpha}}
	=
	\biggl\{
	1-
	\epsilon \delta \fder{}{\dfield}
	\cdot
	\biggl(
		\ep  \cdot \varrho \cdot \fder{}{\dfield}
		-  \dfield 
	\biggr)
	\biggr\}
	e^{-\Stotd[\dfield]}
.
\label{eq:redundant-Stot}
\ee
On account of the total functional derivative on the \rhs, it follows that the partition function is invariant under an infinitesimal shift of the action in the direction of the (integrated) trace of the energy-momentum tensor; in standard parlance, this field is redundant.

It has been appreciated for a long time that every critical fixed-point solution of the ERG equation in fact exists as a line of physically equivalent fixed-points~\cite{Wegner-CS,WegnerInv,TRM-Elements,Fundamentals,RGN}. The exactly marginal, redundant field generates infinitesimal motion along this line:
if the line is parametrized by $b$, then
\be
	\Sintd(b+\delta b) =  \Sintd(b) + \delta b \quasi^{(\D-\delta)} \cdot \quasi^{(\delta)}
.
\ee
For the canonical ERG equation, a generic expression for the entire line of fixed-points can be found in~\cite{Fundamentals} (see also~\cite{OJR-Pohl,OJR-1PI}). Within the ERG formalism, it has been shown that for any (quasi-local) fixed-point for which the exactly marginal, redundant field exists, the value of $\eta$ is isolated~\cite{Fundamentals} (the converse was proven in~\cite{Wegner-CS}). From the perspective of this paper, this property can now be understood as arising for quasi-local theories for which the (trace of the) energy-momentum tensor exists and is non-zero.

\subsection{The Gaussian Fixed-Point}
\label{sec:GFP}

An instructive illustration of many of the concepts discussed above is provided by the Gaussian fixed-point, which describes a free theory for which the fundamental field has scaling dimension $\delta = \delta_0 = (\D-2)/2$.
Before providing the canonical ERG representation of this theory, we will derive the expression for the energy-momentum tensor in the para-Schwinger functional representation. We do this since, for the special case of the Gaussian fixed-point, this representation is, in fact, strictly local and, as such, equivalent to  an unregularized action approach. Consequently, this should provide a familiar setting prior to our exposition of the less conventional ERG approach. Note that the difference between 
$\quasi^{(\D-\delta_0)}$ and $\hat\quasi^{(\D-\delta_0)}$ amounts only to a vacuum term, which we ignore (the same is true in \sect{non-unitary}).

\subsubsection{Para-Schwinger Functional Representation}

In this representation, (and employing canonical normalization) we have
$\quasi^{(\D-\delta_0)} = -\partial^2\dfield$ and $\quasi^{(\delta_0)} = \dfield$, from which we observe that:
\begin{align}
	-\partial^2\dfield \times \partial_\beta \dfield 
&	= 
	-\partial_\alpha
	\bigl(
		 \partial_\alpha \dfield
		\times \partial_\beta \dfield
	\bigr)
	-
	\hf
	\partial_\beta
	\bigl(
		\partial_\lambda \dfield \times \partial_\lambda \dfield
	\bigr)
\nonumber
\\
&	=
	-\partial_\alpha
	\Bigl(
		\partial_\alpha \dfield
		\times \partial_\beta \dfield
		-\hf \delta_{\alpha\beta}
		\partial_\lambda \dfield \times \partial_\lambda \dfield
	\Bigr)
.
\end{align}
Upon comparison with~\eq{divT}, and using a tilde to denote the para-Schwinger functional representation, it is apparent that:
\be
	\tilde{\mathcal{T}}_{\alpha \beta} = \partial_\alpha \dfield \times \partial_\beta \dfield
	-\hf \delta_{\alpha \beta} \partial_\lambda \dfield \times \partial_\lambda \dfield
	+
	\frac{\delta_0}{\D-1}
	\partial_\lambda Z_{\lambda \alpha \beta}
,
\ee
where $Z_{\lambda\alpha\beta}(\dfield)$ satisfies the conditions~\eq{Z_lab} but is thus far undetermined; the factor of $\delta_0/(\D-1)$ has been inserted for convenience. 
Taking the trace and comparing with~\eq{traceT}:
\be
	-\frac{\D-2}{2} \partial_\lambda \dfield \times \partial_\lambda \dfield
	+\frac{\delta_0}{\D-1} \partial_\lambda Z_{\lambda \alpha \alpha}
	=
	\delta_0 \partial^2 \dfield \times \dfield.
\ee
This is solved by taking
\be
	\partial_\lambda Z_{\lambda \alpha \beta} =
	\bigl(
		\delta_{\alpha\beta} \partial^2  - \partial_{\alpha}{\partial_\beta}
	\bigr)
	\hf \dfield^2,
\ee
yielding the standard Gaussian energy-momentum tensor:
\be
	\tilde{\mathcal{T}}_{\alpha \beta} = \partial_\alpha \dfield \times \partial_\beta \dfield
	-\hf \delta_{\alpha \beta} \partial_\lambda \dfield \times \partial_\lambda \dfield
	+
	\frac{\D-2}{4(\D-1)}
	\bigl(
		\delta_{\alpha\beta} \partial^2  - \partial_{\alpha}{\partial_\beta}
	\bigr)
	\dfield^2
.
\label{eq:GaussianEMT}
\ee

Before moving on, it is instructive to consider the expression for the energy-momentum tensor in the Schwinger functional representation. Recalling~\eqs{J-shift}{GreenFunction}, this can be obtained from~\eq{GaussianEMT} simply by making the substitution
\be
	\dfield \rightarrow -\ep_0 \cdot \qsource
.
\ee
It is thus apparent that, in the Schwinger functional representation,
\begin{subequations}
\begin{align}
	\partial_\alpha \mathcal{T}_{\alpha \beta}
	& = 
	- \qsource \times \partial_\beta \ep_0 \cdot \qsource
,
\\
	\mathcal{T}_{\alpha \alpha}
	&=
	-\delta_0 \qsource \times \ep_0 \cdot \qsource
,
\end{align}
\end{subequations}
precisely as we expect for the Ward identities involving the connected correlator of the Gaussian theory.

\subsubsection{ERG Representation}
\label{sec:Gaussian-ERG}

In the ERG formalism, the Gaussian theory exhibits a line of physically equivalent fixed-points, as expected, which terminates in a non-critical fixed-point for which both the Schwinger functional vanishes and the energy-momentum tensor vanish. The Gaussian solution of the canonical ERG equation is (in momentum space):
\be
	\Sintd[\dfield]
	=
	\hf
	\int_p 
	\dfield(p)
	\frac{b p^2}{1-b \cutoff(p^2)} \dfield(-p),
	\qquad
	\Stotd[\dfield]
	=
	\hf
	\int_p 
	\dfield(p)
	\frac{p^2 \cutoff^{-1}(p^2)}{1-b \cutoff(p^2)} \dfield(-p)
,
\ee
where, in accord with convention, $\cutoff(0) = 1$ and $\int_p \equiv (2\pi)^{-\D}\Int{p}$. It is thus apparent that the Gaussian fixed point exists as a line for $-\infty <b < 1$. At $b=1$, the action is still a fixed-point in the sense of solving the ERG equation. However, Taylor expanding $\cutoff(p^2) = 1 + \order{p^2}$, it is clear that the action does not describe a theory with long-range order: this theory is non-critical. As emphasised by Wegner, such theories support only redundant fields and, as we will see below, both the energy-momentum tensor and the correlation functions vanish.
Note that, for $b>1$, the coefficient of $\dfield(p) p^2 \dfield(-p)$ turns negative, which manifests itself as a loss of positivity of the two-point correlation function, as will also be seen below.

Defining
\be
	\overline{G}
	=
	(1-b\cutoff)^{-1},
\ee
the Gaussian solution can be rewritten:
\be
	\Stotd[\dfield] = 
	\hf 
	\dfield
	\cdot \ep^{-1} \cdot \overline{G} \cdot \dfield
.
\ee
Substituting into~\eqs{quasi_delta-local}{quasi_d-delta-local}, it can readily be confirmed that
\begin{subequations}
\begin{align}
	\quasi^{(\delta)}_{\mathrm{loc}}
	&
	= 
	(1-b)
	\overline{G} \cdot \dfield
,
\label{eq:quasi_delta_0}
\\
	\quasi^{(\D-\delta)}_{\mathrm{loc}} & 
	=  
	 \dfield \cdot \overline{G} \cdot  \ep^{-1}_0
.
\label{eq:quasi_d-delta_0}
\end{align}
\end{subequations}
The full energy-momentum tensor can be obtained from~\eq{GaussianEMT} by making the substitution
\be
	\dfield(x) \rightarrow \sqrt{1-b} \Int{y} \overline{G}\bigl((x-y)^2 \bigr) \dfield(y)
.
\ee
Furthermore, as shown explicitly in~\cite{Fundamentals}, the Schwinger functional for the Gaussian theory is given by
\be
	\dual[\qsource] = \frac{1-b}{2} \int_p \qsource(p) \frac{1}{p^2} \qsource(-p)
.
\ee
Thus it is apparent that, at the point the theory turns non-critical, both the correlation functions and the energy-momentum tensor vanish. For $b>1$, positivity of the two-point function is violated.

\subsection{A Non-Unitary Example}
\label{sec:non-unitary}

Having observed the successful construction of the energy-momentum tensor in the simplest local, unitary theory, in this section we will provide a simple non-unitary example where the construction breaks down---at least in $\D=2$. (For a pedagogical exposition of various aspects of unitarity, see~\cite{Intriligator-Unparticles}.) Specifically, we will consider a free theory for which $\eta = -2$. 

In momentum space, it is a simple matter to show that~\cite{Fundamentals}
\be
	\Sintd = \hf\int_p \dfield(p) \dfield(-p) \frac{bp^2 - p^4 \cutoff^{-1}(1+\varrho)}{\varrho p^2 - b\cutoff},
\ee
where $b$ parametrizes the line of equivalent fixed-points and $\varrho$ is given by~\eq{rho}. Note that the two-point function for the full action, \cf~\eq{TotalAction}, starts at $\order{p^4}$.
It follows that
\begin{subequations}
\begin{align}
	\quasi^{(\delta)}(p) &= -\frac{b}{p^2 \varrho - b\cutoff} \dfield(p),
\\
	\quasi^{(\D-\delta)}(p) & = -\frac{p^4}{\varrho p^2 - b\cutoff} \dfield(p).
\end{align}
\end{subequations}
Taking $b=-1$ (to attain canonical normalization) let us define
\be
	\Phi = (p^2 \varrho + \cutoff)^{-1} \cdot \dfield
,
\ee
giving $\quasi^{(\D-\delta)} = -\partial^4 \Phi$ and $\quasi^{(\delta)} = \Phi$. The attempted construction of the energy-momentum tensor precedes in a similar vein to before. Observe that
\be
	\partial_\alpha \Phi \times \partial^4 \Phi
	=
	\partial_\rho
	\bigl(
		\partial_\alpha \Phi \times \partial_\rho \partial^2 \Phi
		+ \partial_\rho \Phi \times \partial_\alpha \partial^2 \Phi
	\bigr)
	-
	\hf \partial_\alpha
	\bigl(
		\partial^2 \Phi \times \partial^2 \Phi
		+ 2 \partial_\rho \Phi \times \partial_\rho \partial^2 \Phi
	\bigr)
.
\ee
Therefore, \eq{divT} implies
\be
	T_{\alpha \beta} = 
	\partial_\alpha \Phi \times \partial_\beta \partial^2 \Phi
	+\partial_\beta \Phi \times \partial_\alpha \partial^2 \Phi
	-\hf \delta_{\alpha\beta}
	\bigl(
		\partial^2 \Phi \times \partial^2 \Phi
		+ 2 \partial_\rho \Phi \times \partial_\rho \partial^2 \Phi
	\bigr)
	+ \partial_\lambda Z_{\lambda \alpha \beta}
.
\ee
Taking the trace yields
\be
	{T}_{\alpha \alpha}
	=
	(2-\D) \partial_\alpha \Phi \times \partial_\alpha \partial^2 \Phi
	-\frac{\D}{2} \partial^2 \Phi \times \partial^2 \Phi + \partial_\lambda Z_{\lambda \alpha \alpha}
\ee
However, from~\eq{traceT}, we know that the trace of the energy-momentum tensor is to be equated with $-\delta \quasi^{(\D-\delta)} \times \quasi^{(\delta)}$, which in this case amounts to 
setting
\be
	(2-\D) \partial_\alpha \Phi \times \partial_\alpha \partial^2 \Phi
	-\frac{\D}{2} \partial^2 \Phi \times \partial^2 \Phi + \partial_\lambda Z_{\lambda \alpha \alpha}
	= \frac{\D-4}{2} \Phi \times \partial^4 \Phi
.
\ee
This simplifies to
\be
	2 \partial_\alpha 
	\bigl(
		\Phi \times \partial_\alpha \partial^2 \Phi
	\bigr)
	-
	\frac{\D}{2} \partial^2
	\bigl(
		\Phi \times \partial^2 \Phi
	\bigr)
	+
	\partial_\lambda Z_{\lambda \alpha \alpha} = 0
.
\label{eq:Z_laa}
\ee
Comparing with~\eq{Pol-condition-d=2} we see that, in $\D=2$, the presence of the first term prevents the construction of a quasi-local $\partial_\lambda Z_{\lambda \alpha \beta}$. 

Notice that 
\be
	\Phi \times \partial_\alpha \partial^2 \Phi
	=
	\partial_\sigma
	\Bigl(
		\Phi \times \partial_\alpha \partial_\sigma \Phi
		-
		\hf
		\delta_{\sigma \alpha} \partial_\lambda \Phi \times \partial_\lambda \Phi
	\Bigr)
.
\ee
In $\D>2$, improvement of the energy-momentum tensor along the lines of~\eq{Pol-condition-d>2} is thus possible. For $\D=2$ the energy-momentum tensor cannot be improved. Recalling the discussion around~\eq{NoConformal}, the $\eta = -2$ free theory thus provides an example where, in the quasi-local representation, the action is apparently conformally invariant but, nevertheless, the full quantum theory is not. It is easy to see that $\dual[\qsource]$ does not exist, since the propagator $\sim 1/p^4$.

To conclude, note that for $\D>2$ the energy-momentum tensor can be improved by taking%
\footnote{I am very grateful to Hidenori Sonoda for supplying me with this solution.}:
\be
	Z_{\lambda \alpha \beta} = -\partial_\mu Y_{\alpha \lambda \beta \mu},
\ee 
where
\begin{multline}
	Y_{\alpha \lambda \beta \mu}
	=
	\bigl(
		\delta_{\alpha\beta} \delta_{\lambda \mu} - \delta_{\alpha \mu} \delta_{\lambda \beta}
	\bigr)
	\frac{1}{\D-1}
	\biggl(
		\frac{\D}{\D-2} \partial_\gamma \Phi \times \partial_\gamma \Phi
		- \frac{\D-4}{2} \Phi \times \partial^2 \Phi
	\biggr)
\\
	-\frac{2}{\D-2}
	\bigl(
		\delta_{\alpha \beta} \partial_\lambda \Phi \times \partial_\mu \Phi
		+ \delta_{\lambda \mu} \partial_\alpha \Phi \times \partial_\beta \Phi
		- \delta_{\lambda \beta} \partial_\alpha \Phi \times \partial_\mu \Phi
		-\delta_{\alpha \mu} \partial_\lambda \Phi \times \partial_\beta \Phi
	\bigr)
.
\end{multline}
By virtue of the symmetries
\be
	Y_{\alpha \lambda \beta \mu} = - Y_{\lambda \alpha \beta \mu} = Y_{\beta \mu \alpha \lambda},
\ee
it is apparent that, as required,
\be
	\partial_\alpha \partial_\lambda Z_{\lambda \alpha \beta} = 0,
	\qquad
	\partial_\lambda Z_{\lambda \alpha \beta}  = \partial_\lambda Z_{\lambda \beta \alpha} 
.
\ee
Finally, it can be checked that, upon taking the trace, we recover~\eq{Z_laa}.

\section{Conclusions}

The essence of the philosophy advocated in this paper is a conservative one. At its heart is the desire to view, as far as possible, QFT as fundamental. (Whether or not this ultimately turns out to be the case is beside the point; the goal is partly to see how far one can go by pursuing this agenda.) Taking this seriously, we are driven to look for theories which make sense down to arbitrarily short distances (\ie\ theories exhibiting non-perturbative renormalizability). Wilsonian renormalization teaches us that a sufficient condition for this is scale-invariant behaviour in the deep UV, suggesting that
we investigate either fully scale-invariant theories or relevant/marginally relevant deformations, thereof.  

Our exclusive focus has been on theories exhibiting invariance under the full conformal group.
In this context, there are two largely disjoint approaches: one based on exploring the constraints implied by conformal invariance on the correlation functions and the other a path integral approach built upon a quasi-local action. True to our philosophy, the former is viewed as more primitive due to its inherently quantum field-theoretic nature whereas, through the action, the latter manifests its classical heritage. Ideally, then, what we would like is to be able to start with an approach based on the correlation functions and to show how an action-based description emerges.

This paper largely shows how to achieve this. Starting from the statements of conformal covariance of the correlation functions, the first step is to wrap these up into functionals of sources (accepting a degree of formality in this step). Associated with this is a functional representation of the conformal algebra. This forms the basis for constructing more elaborate representations, involving auxiliary functionals. These auxiliary functionals satisfy consistency conditions, and our development culminated with a representation in which the condition corresponding to dilatation invariance is nothing but the fixed-point version of the canonical ERG equation of~\cite{Ball}. The explicit form of the partner encoding special conformal invariance is a new result of this paper.

Nevertheless, it must be acknowledged that we were guided towards this representation because we knew what we were looking for. This, in of itself, is not an issue. More pressing is that, coming from the path integral  perspective, it is expected that all physically acceptable solutions to ERG equations correspond to quasi-local actions~\cite{Wilson,Fundamentals}. The step that is missing in this paper is to show that ERG representations of CFTs necessarily have a Wilsonian effective action that is, indeed, quasi-local. Or, to put it another way, of all the possible representations of the conformal algebra, what makes the ERG representation so special?
Suggestively, as alluded to in \sect{qlrep}, if the ERG representation of a CFT is quasi-local, then the energy-momentum tensor is amongst the spectrum of conformal primaries. 
Indeed, the ERG and the energy-momentum tensor share an intimate relationship revealed in this paper: lines of physically equivalent fixed-points are generated by the trace of the energy-momentum tensor.

Therefore, referring to the questions posed in the introduction, the state of affairs is as follows. A concrete recipe has been provided for encoding conformal dynamics in an object recognizable as the Wilsonian effective action. This assumes~\eq{dualFromS} which, given the choice~\eq{op}, amounts to an assumption of the existence of a path integral. As noted, in $\D=2$ it may be necessary to work in finite volume. Plausibly, the rather formal process presented  will work for theories possessing an energy-momentum tensor, in which case we expect the Wilsonian representation to furnish a quasi-local formulation of the theory in question. It is clearly desirable, however, to place all of this on a more rigorous footing.

Beyond this matter, it is worth posing the question as to whether there may exist theories supporting representations of the conformal algebra for which the constraint which picks out physically acceptable theories is entirely different from quasi-locality, but equally powerful.  Besides exploring this theme further, several avenues of future research suggest themselves. 
Transcribing our approach to supersymmetric theories should be largely a matter of working with the appropriate potential superfields, as in~\cite{Susy-Chiral}, and being mindful that the conformal algebra is enhanced to be superconformal. A controlled environment in which to further explore the CFT/ERG link is in $\D=2$, where is may be profitable to investigate functional representations of the entire Virasoro algebra. As indicated earlier, a rigorous treatment in $\D=2$ may entail a careful finite-volume treatment. Gauge theories present special problems~\cite{Fundamentals,JMP-Review} and it is my belief that appropriately extending the ideas of this paper will require new ideas.

Furthermore, it is desirable to extend the scope of the analysis to include scale-dependent theories. It is anticipated that the renormalizability (or otherwise) of such theories is tied up with the renormalization of composite operators. Since, for irrelevant/marginally irrelevant perturbations, renormalizability is lost this suggests that, in order to properly define the fixed-point Schwinger functional involving the corresponding sources, some form of point splitting should be performed on the composite operators, to improve their UV behaviour. It may be that this engenders a natural way to uncover the operator product expansion within the ERG formalism, raising the hope of making concrete links between the ideas of this paper and recent developments in the application of the conformal bootstrap. 

Finally, it is worth considering the question as to whether theories exist in which there are multiple, distinct quasi-local representations. This leads naturally to the subject of dualities and it is hoped that the ideas of this paper and the fresh perspective it gives on the nature and origins of Wilsonian renormalization will offer new insights in this area.

\begin{acknowledgments}
This paper is dedicated to the memory of my friend, Francis Dolan, who died, tragically, in 2011. It is gratifying that I have been able to honour him with work which substantially overlaps with his research interests and also that some of the inspiration came from a long dialogue with his mentor and collaborator, Hugh Osborn. In addition, I am indebted to Hugh for numerous perceptive comments on various drafts of the manuscript and for bringing to my attention gaps in my knowledge and holes in my logic. I would like to thank Yu Nakayama and Hidenori Sonoda for insightful correspondence following the appearance of the first and third versions on the arXiv, respectively.

I am firmly of the conviction that the psychological brutality of the post-doctoral system played a strong underlying role in Francis' death. I would like to take this opportunity, should anyone be listening, to urge those within academia in roles of leadership to do far more to protect members of the community suffering from mental health problems, particularly during the most vulnerable stages of their careers.
\end{acknowledgments}

\end{document}